\documentclass[5p,twocolumn,times]{elsarticle}

\usepackage{lipsum}
\usepackage{lineno,hyperref}
\modulolinenumbers[5]
\usepackage[pdftex]{color}
\usepackage[font=footnotesize,labelfont=bf]{caption}
\usepackage[font=footnotesize,labelfont=bf]{subcaption}
\journal{Journal of \LaTeX\ Templates}

\usepackage{amssymb}

\biboptions{compress}

\usepackage[figuresright]{rotating}

\begin{document}

\begin{frontmatter}

\title{Mobility restrictions in response to local epidemic outbreaks in rock-paper-scissors models}

\address[1]{School of Science and Technology, Federal University of Rio Grande do Norte\\
59072-970, P.O. Box 1524, Natal, RN, Brazil}
\address[2]{Institute for Biodiversity and Ecosystem
Dynamics, University of Amsterdam, Science Park 904, 1098 XH
Amsterdam, The Netherlands}

\author[1,2]{J. Menezes}

\begin{abstract}
We study a three-species cyclic model whose organisms are vulnerable to contamination with an infectious disease which propagates person-to-person. 
We consider that individuals of one species perform an evolutionary self-preservation strategy by reducing the mobility rate to minimise infection risk whenever an epidemic outbreak reaches the neighbourhood. Running stochastic simulations, we quantify the changes in spatial patterns induced by unevenness in the cyclic game introduced by the mobility restriction strategy of organisms of one out of the species. Our findings show that variations in disease virulence impact the benefits of dispersal limitation reaction, with the relative reduction of the organisms' infection risk accentuating in surges of less contagious or deadlier diseases. The effectiveness of the mobility restriction tactic depends on the tolerable fraction of infected neighbours used as a trigger of the defensive strategy and the deceleration level. If each organism promptly reacts to the arrival of the first viral vectors in its surroundings with strict mobility reduction,  contamination risk decreases significantly. Our conclusions may help biologists understand the impact of evolutionary defensive strategies in ecosystems during an epidemic.
\end{abstract}

\end{frontmatter}

\section{Introduction}
\label{sec1}
Experiments with three strains of \textit{Escherichia coli} revealed that the coexistence results from the nonhierarchical cyclic dominance, described by the rock-paper-scissors game rules \cite{bacteria}. Nevertheless, the scientists observed that
species persistence was possible only if the selection interactions occurred locally, leading to the formation of departed spatial domains \cite{Coli,Allelopathy}.
This means that cyclic interactions are insufficient to maintain biodiversity, which reveals that
organisms' mobility plays a central role in biodiversity maintenance
 \cite{ecology,Nature-bio,mobilia2,mobilia3}. This phenomenon has also been found in other biological systems, like Californian coral reef invertebrates and lizards in the inner Coast Range of California, which confirms the central role space plays in promoting biodiversity in cyclic game systems \cite{lizards,coral,mobiliahigh}.

Animals' mobility has also been investigated in behavioural ecology as an adaptive response to environmental changes threatening organisms' survival \cite{foraging,butterfly,BUCHHOLZ2007401}. As the species evolves, organisms learn to scan the environment and interpret the signals captured from the neighbourhood; thus, adapting their movement to avoid hostile regions and reach comfort zones, where personal fitness is maximised
\cite{doi:10.1002/ece3.4446,adaptive1,adaptive2,Dispersal,BENHAMOU1989375,Moura,anti2,MENEZES2022101606,TENORIO2022112430,Menezes_2022}. Learning from animals' adaptive movement,
engineers have created sophisticated tools to improve robots that imitate animals' response to environmental changes \cite{animats}.

If a lethal disease spreads through the system, controlling individuals' mobility has been shown an effective strategy for minimising the risk of contamination with viral illness transmitted person-to-person \cite{social1,disease4,tanimoto}. This has been proved, for example, by the individual and collective benefits of the successful social distancing rules worldwide implemented to mitigate the effects of pandemics \cite{socialdist,soc,doi:10.1126/science.abc8881}. The effectiveness of the dispersal restriction response to local disease surges has been quantified by the reduction in the number of infected organisms and, consequently, the minimisation of the social impact of epidemics on communities \cite{mr0,mr1,mr2,10.1371/journal.pone.0254403}. 
Adjusting the protective measures when virus mutation changes the disease virulence is vital to ensure the effectiveness, maximising the benefits of the behavioural strategy \cite{CAPAROGLU2021111246}. This has been observed in the spatial organisation plasticity resulting from adapting the mobility restrictions to improve the profit of the disease mitigation strategy faced alterations in the transmission and mortality rates \cite{plasticity2,Gene,mutating1,mutate2,plasticity1}.

Recently, it has been shown that the mobility limitation tactic
reduces the organisms' disease contamination risk in the spatial rock-paper-scissors model \cite{plasticity}. This changes the spatial patterns, with organisms of the same species occupying regions whose characteristic length scale varies with disease infection and mortality rates. However, the model addressed in Ref.~\cite{plasticity} focused on a globally designed dispersal limitation, equally imposed on all organisms, independent of the existence of a disease outbreak affecting the organisms' neighbourhood. This means that, although unveiling essential details of the complexity of disease transmission mitigation achieved by a global mobility restriction rule, the model does not 
consider the heterogeneity of the spatial distribution of infected organisms, causing local disease outbreaks with distinct intensities.

In this work, instead of assuming a fixed global strategy for all organisms in the system, as in Ref.~\cite{plasticity},
we study a mobility restriction response triggered autonomously by each individual whenever perceiving a disease outbreak in its vicinity. Therefore, if a given organism is not imminent to be contaminated, it continues moving with maximum mobility, advancing on the lattice to conquer territory, allowing population growth \cite{mobilia2}. We consider that the behavioural strategy results from an evolutionary process that gives the ability to individuals of one out of the species to perform three steps: i) scanning the environment and recognise which neighbours are sick; ii) concluding if the local density of viral vectors exceeds a tolerable threshold; iii) reducing its velocity to minimise the chances of being infected by a sick neighbour. Both healthy and sick organisms limit their movement during a local epidemic outbreak. This aims to minimise the number of organisms becoming sick and the probability of infected individuals passing the virus to their conspecifics.

We aim to understand how the mobility limitation response performed by individuals of one species unbalances the 
spatial rock-paper-scissors game during an epidemic. We also objective to comprehend how the choice of the tolerable threshold of the local density of sick individuals influences the success of the behavioural strategy. For this purpose, we perform thousands of simulations considering many levels of mobility restriction triggers and slowness factors. Moreover, we aim to quantify the impact of eventual pathogen mutations altering the disease virulence - transmissibility and mortality - on the benefits of the defensive strategy.

We simulate the stochastic version of the rock-paper-scissors model using the May-Leonard implementation, where the total number of individuals is not conserved \cite{PhysRevE.97.032415,Avelino-PRE-86-036112,BAZEIA2022126547}.
Our epidemic model considers that infected organisms act as viral vectors, with no healthy individual becoming immune when cured of the disease \cite{plasticity,combination, jcomp,eplsick,rps-epidemy,epidemic-graphs,germen}. Because of the unevenness in the cyclic game introduced by the ability of organisms of one out of the species to perform the behavioural self-preservation strategy, the species dominance depends on the local intensities of the epidemic surges inducing individuals to decelerate \cite{unevenmobility,uneven,PedroWeak,unevenpark1,howlocal,MENEZES2023113290}.

\begin{figure}
\centering
\includegraphics[width=50mm]{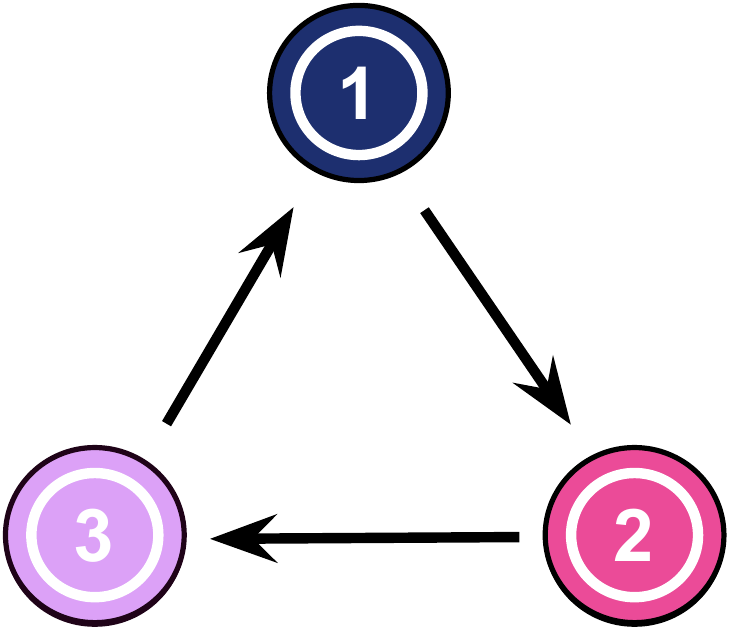}
\caption{Illustration of the spatial rock-paper-scissors game. 
The arrows indicate the species' cyclic dominance, with organisms of species $i$ eliminating individuals of species $i-1$. Throughout this work, blue, pink, and purple stand for organisms of species $1$, $2$, and $3$, respectively.}
	\label{fig1}
\end{figure}

The outline of this paper is as follows: in Sec.~\ref{sec2}, we introduce the model and explain how the simulations are implemented. The pattern formation process and the spatial distribution of disease outbreaks are investigated in 
Sec.~\ref{sec3}. The autocorrelation function and the characteristic length scale of the typical spatial domains occupied by each species are calculated in Sec.~\ref{sec4}. In Sec.~\ref{sec5}, we compute the impact of disease transmission and mortality rates on organisms' infection risk and species densities. The mobility restriction trigger and slowness factor  
are studied in Sec.~\ref{sec6} and ~\ref{sec7}.
Finally, our conclusions and discussion appear in Sec.~\ref{sec8}.


\begin{figure}
 \centering
        \begin{subfigure}{.2\textwidth}
        \centering
        \includegraphics[width=36mm]{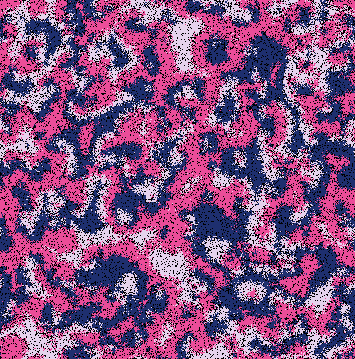}
        \caption{}\label{fig2a}
    \end{subfigure}
   \begin{subfigure}{.2\textwidth}
        \centering
        \includegraphics[width=42mm]{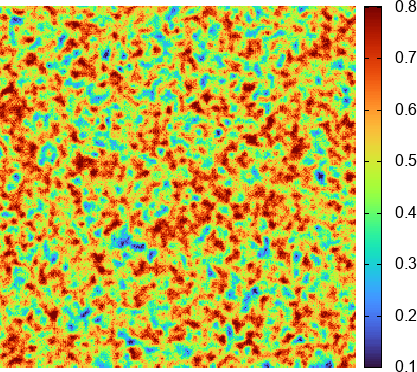}
        \caption{}\label{fig2b}
    \end{subfigure} 
           \begin{subfigure}{.2\textwidth}
        \centering
        \includegraphics[width=36mm]{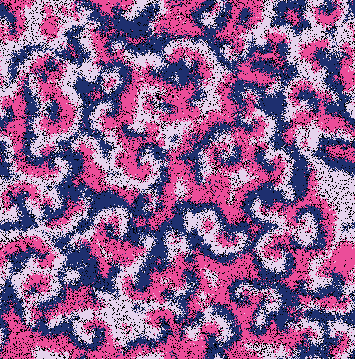}
        \caption{}\label{fig2c}
    \end{subfigure}
            \begin{subfigure}{.2\textwidth}
        \centering
        \includegraphics[width=42mm]{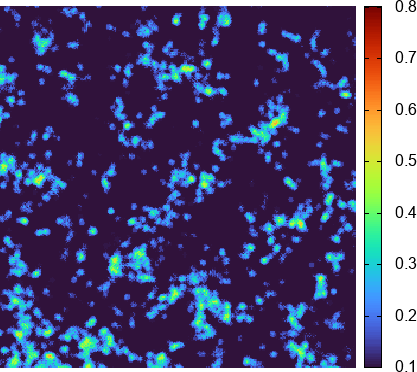}
        \caption{}\label{fig2d}
    \end{subfigure}
   \begin{subfigure}{.2\textwidth}
        \centering
        \includegraphics[width=36mm]{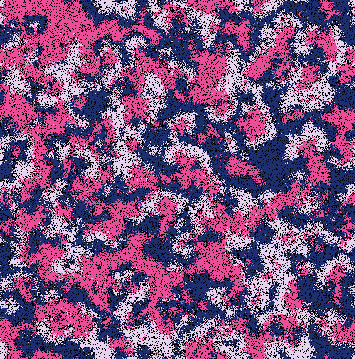}
        \caption{}\label{fig2e}
    \end{subfigure} 
           \begin{subfigure}{.2\textwidth}
        \centering
        \includegraphics[width=42mm]{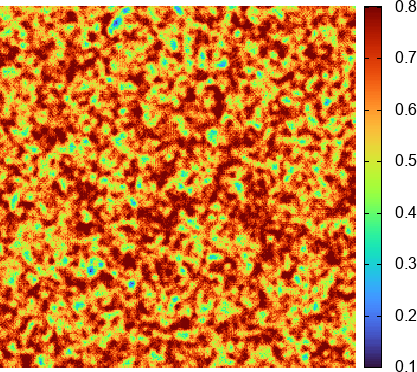}
        \caption{}\label{fig2f}
    \end{subfigure} 
\caption{Spatial patterns and local epidemic outbreaks in the rock-paper-scissors model. 
Figures \ref{fig2a}, \ref{fig2c}, and \ref{fig2e} show the final organisms' spatial organisation for Simulations A (https://youtu.be/mSsL0XfMtlI), B (https://youtu.be/Iwy25cR9bsM), and C (https://youtu.be/2a1pieelQrc). Blue, pink, and purple dots depict individuals of species $1$, $2$, and $3$, respectively; black dots indicate the empty spaces. The colour bar in Figures \ref{fig2b}, \ref{fig2d}, and \ref{fig2f} shows the final spatial distribution of the local density of sick individuals in Simulations A (https://youtu.be/i7q9RsI0ZW0), B (https://youtu.be/T1vnJ2aRBgQ), and C (https://youtu.be/ll17ZnGh1cc).}
  \label{fig2}
\end{figure}
\begin{figure}[h]
\centering
       \begin{subfigure}{.48\textwidth}
        \centering
        \includegraphics[width=85mm]{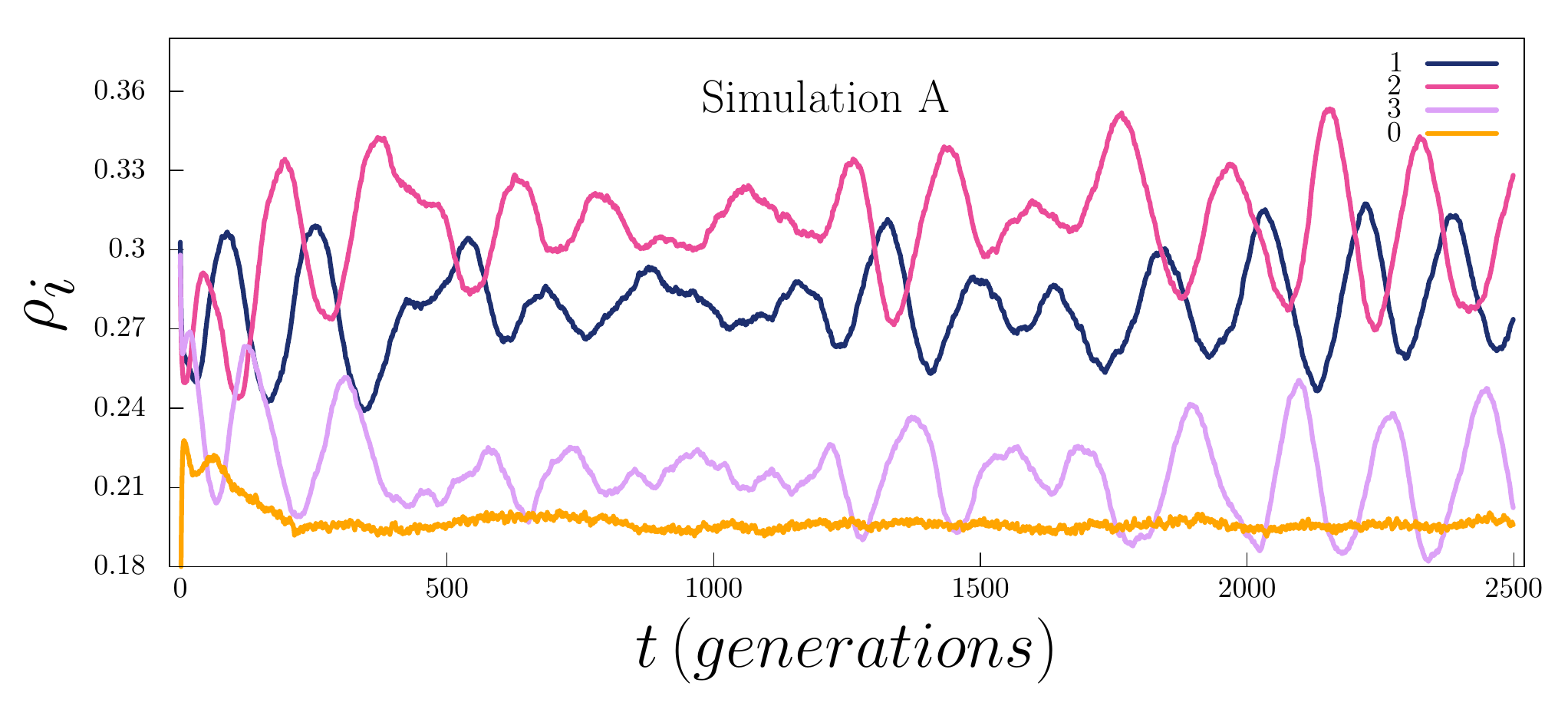}
        \caption{}\label{fig3a}
    \end{subfigure}\\
           \begin{subfigure}{.48\textwidth}
        \centering
        \includegraphics[width=85mm]{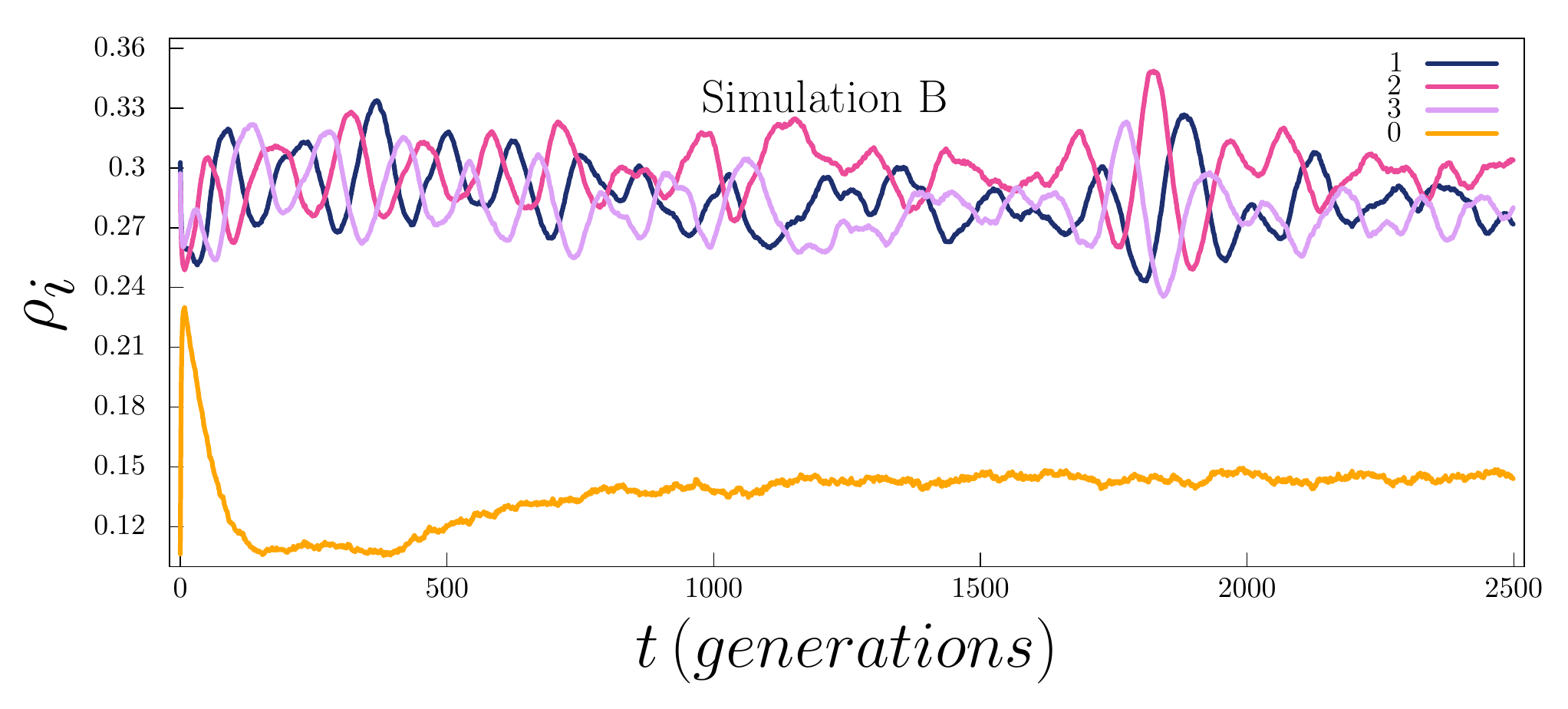}
        \caption{}\label{fig3b}
    \end{subfigure}\\
           \begin{subfigure}{.48\textwidth}
        \centering
        \includegraphics[width=85mm]{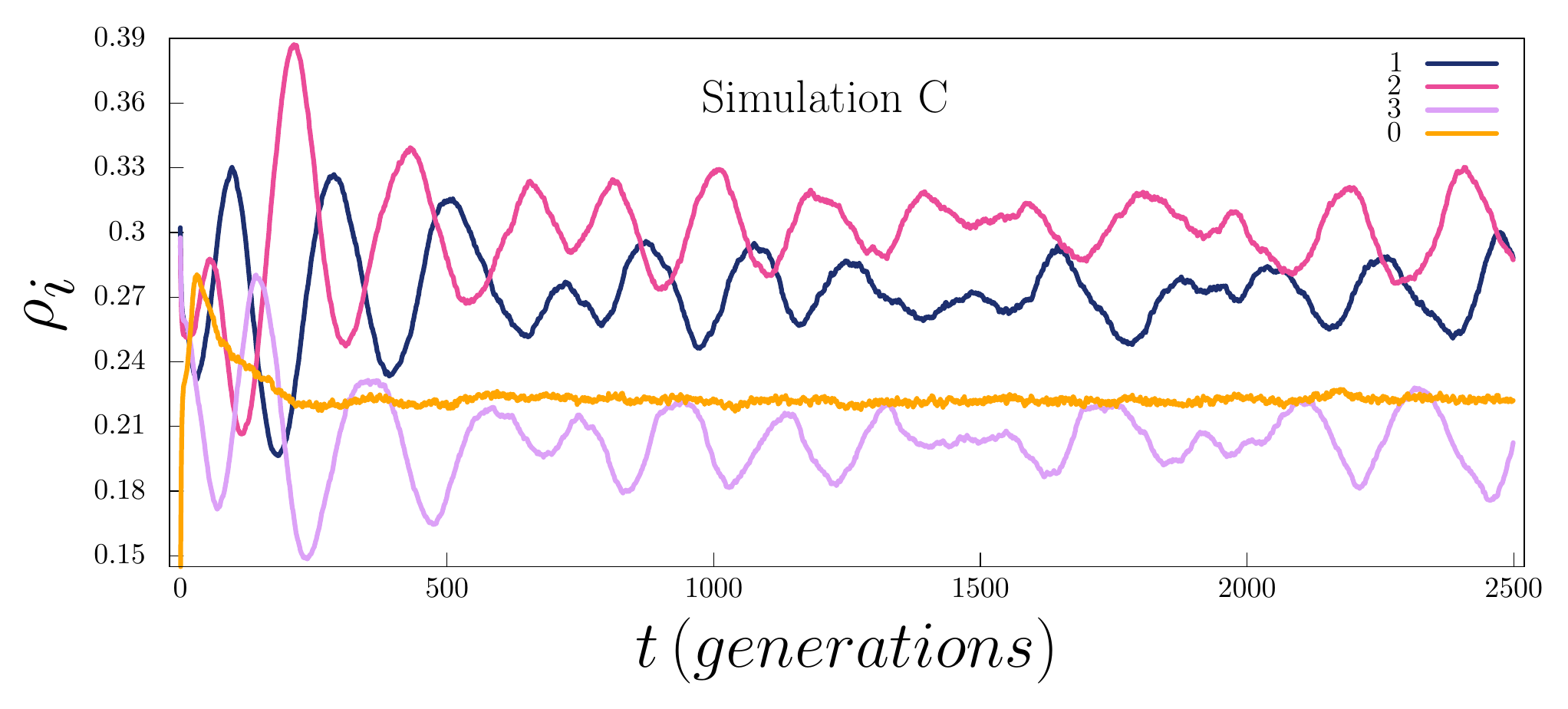}
        \caption{}\label{fig3c}
    \end{subfigure}
\caption{Dynamics of species densities for various disease virulence. Figures ~\ref{fig3a}, 
~\ref{fig3b}, and ~\ref{fig3c} shows the temporal dependence of the densities of organisms of species $1$ (blue line), $2$ (pink line), and $3$ (purple lines) in Simulations A, B, and C, respectively. The orange lines depict the density of empty spaces, $\rho_0$.}
\label{fig3}
\end{figure}

\begin{figure*}[h]
	\centering
	  \begin{subfigure}{.19\textwidth}
        \centering
        \includegraphics[width=34mm]{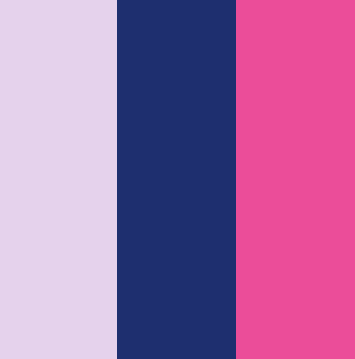}
        \caption{}\label{fig4a}
    \end{subfigure} 
    \begin{subfigure}{.19\textwidth}
        \centering
        \includegraphics[width=34mm]{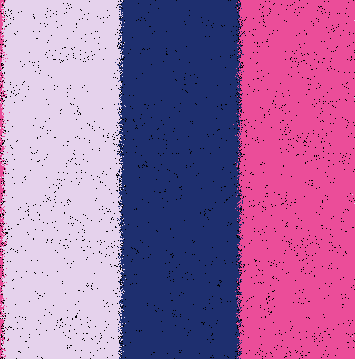}
        \caption{}\label{fig4b}
    \end{subfigure} %
   \begin{subfigure}{.19\textwidth}
        \centering
        \includegraphics[width=34mm]{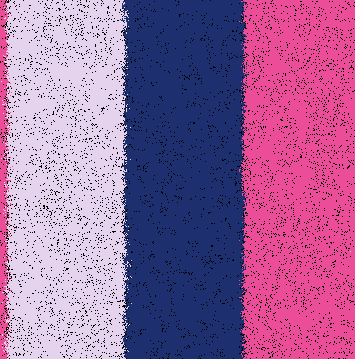}
        \caption{}\label{fig4c}
    \end{subfigure} 
            \begin{subfigure}{.19\textwidth}
        \centering
        \includegraphics[width=34mm]{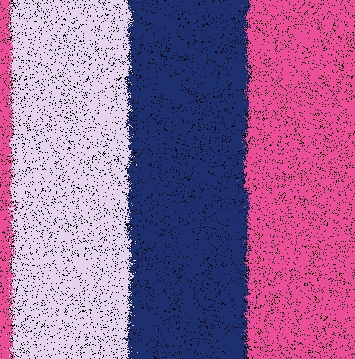}
        \caption{}\label{fig4d}
    \end{subfigure} 
           \begin{subfigure}{.19\textwidth}
        \centering
        \includegraphics[width=34mm]{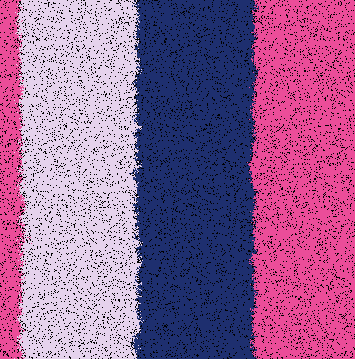}
        \caption{}\label{fig4e}
    \end{subfigure} \\
                \begin{subfigure}{.19\textwidth}
        \centering
        \includegraphics[width=34mm]{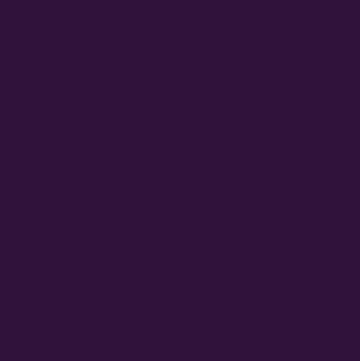}
        \caption{}\label{fig4f}
    \end{subfigure} %
   \begin{subfigure}{.19\textwidth}
        \centering
        \includegraphics[width=34mm]{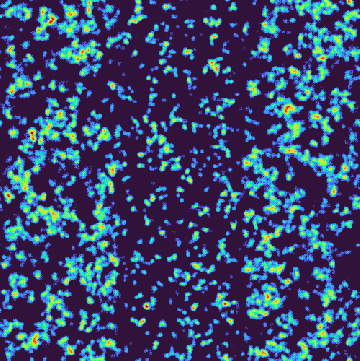}
        \caption{}\label{fig4g}
    \end{subfigure} 
            \begin{subfigure}{.19\textwidth}
        \centering
        \includegraphics[width=34mm]{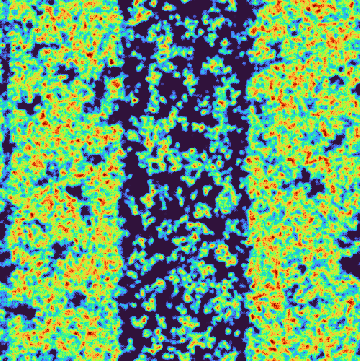}
        \caption{}\label{fig4h}
    \end{subfigure} 
           \begin{subfigure}{.19\textwidth}
        \centering
        \includegraphics[width=34mm]{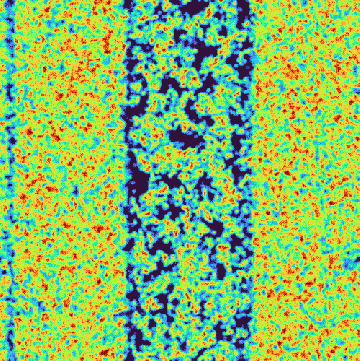}
        \caption{}\label{fig4i}
    \end{subfigure} 
   \begin{subfigure}{.19\textwidth}
        \centering
        \includegraphics[width=34mm]{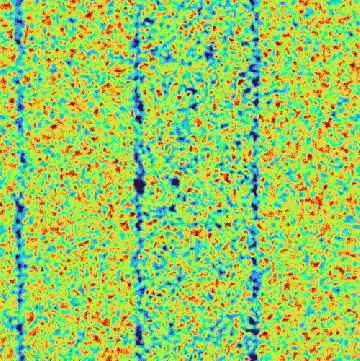}
        \caption{}\label{fig4j}
            \end{subfigure}
 \caption{Snapshots of the initial stage of Simulation D, starting from the prepared initial conditions in Fig.~\ref{fig4a}. 
The organisms' spatial organisation at $t=20$, $t=40$, $t=60$, and $t=100$ are showed in Figs.~\ref{fig4b}, ~\ref{fig4c}, ~\ref{fig4d}, and ~\ref{fig4e}, respectively. The colours follow the scheme in Fig~\ref{fig1}; empty spaces as depicted by black dots. Figure ~\ref{fig4f}, ~\ref{fig4g}, ~\ref{fig4h}, ~\ref{fig4i} and ~\ref{fig4j} show the local densities of sick individuals in Figs.~\ref{fig4a}, ~\ref{fig4b}, ~\ref{fig4c}, ~\ref{fig4d} and ~\ref{fig4e}, respectively, according to the colour bars in Fig. \ref{fig2}. The simulation ran for $\kappa=1.0$ and $\mu=0.2$.}
  \label{fig4}
\end{figure*}

\begin{figure*}
	\centering
    \begin{subfigure}{.19\textwidth}
        \centering
        \includegraphics[width=34mm]{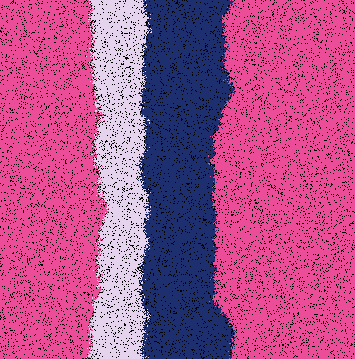}
        \caption{}\label{fig5a}
    \end{subfigure} %
   \begin{subfigure}{.19\textwidth}
        \centering
        \includegraphics[width=34mm]{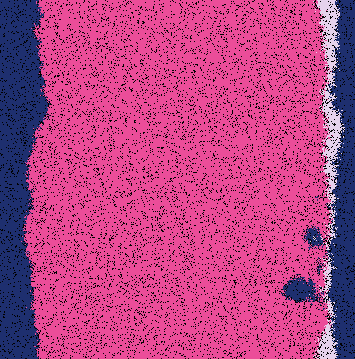}
        \caption{}\label{fig5b}
    \end{subfigure} 
            \begin{subfigure}{.19\textwidth}
        \centering
        \includegraphics[width=34mm]{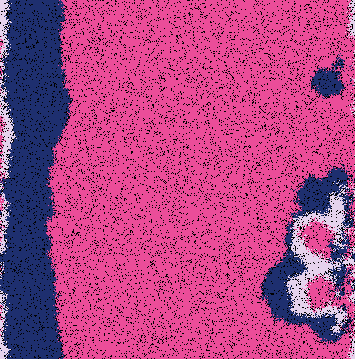}
        \caption{}\label{fig5c}
    \end{subfigure} 
           \begin{subfigure}{.19\textwidth}
        \centering
        \includegraphics[width=34mm]{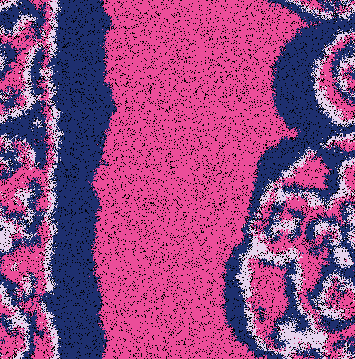}
        \caption{}\label{fig5d}
    \end{subfigure} 
   \begin{subfigure}{.19\textwidth}
        \centering
        \includegraphics[width=34mm]{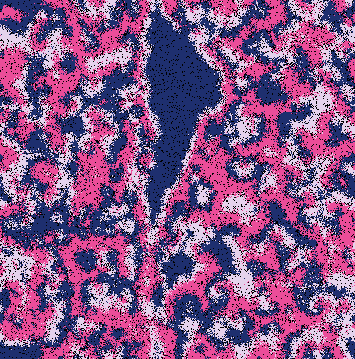}
        \caption{}\label{fig5e}
            \end{subfigure}\\
                \begin{subfigure}{.19\textwidth}
        \centering
        \includegraphics[width=34mm]{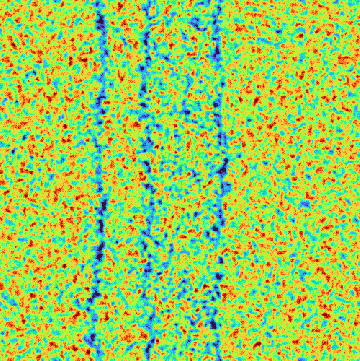}
        \caption{}\label{fig5f}
    \end{subfigure} %
   \begin{subfigure}{.19\textwidth}
        \centering
        \includegraphics[width=34mm]{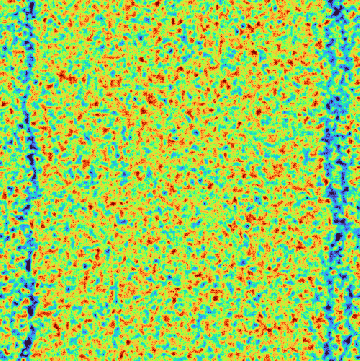}
        \caption{}\label{fig5g}
    \end{subfigure} 
            \begin{subfigure}{.19\textwidth}
        \centering
        \includegraphics[width=34mm]{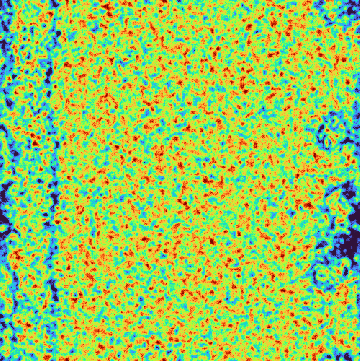}
        \caption{}\label{fig5h}
    \end{subfigure} 
           \begin{subfigure}{.19\textwidth}
        \centering
        \includegraphics[width=34mm]{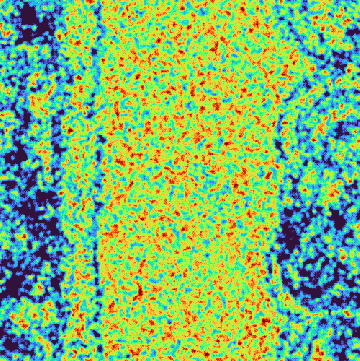}
        \caption{}\label{fig5i}
    \end{subfigure} 
   \begin{subfigure}{.19\textwidth}
        \centering
        \includegraphics[width=34mm]{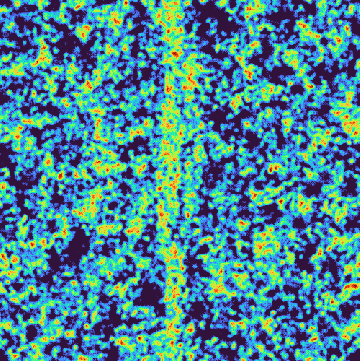}
        \caption{}\label{fig5j}
            \end{subfigure}
 \caption{Snapshots of the final stage of Simulation D, starting from the prepared initial conditions in Fig.~\ref{fig5a}. 
Figures ~\ref{fig5a}, ~\ref{fig5b}, ~\ref{fig5c}, ~\ref{fig5d}, and ~\ref{fig4e} show the individuals' spatial distribution 
at $t=2000$, $t=3000$, $t=3140$, $t=3400$, and $t=3920$, respectively. The respective spatial distribution of the local density of sick organisms appears in 
Figs. ~\ref{fig5f}, ~\ref{fig5g}, ~\ref{fig5h}, ~\ref{fig5i}, and ~\ref{fig5j}. Videos https://youtu.be/WTa-QUa3NFw and https://youtu.be/VZkpNzW2-cs show the dynamics of the spatial patterns and 
local disease outbreaks during the entire simulation.}
  \label{fig5}
\end{figure*}


\section{The stochastic model}
\label{sec2}

We study the spatial version of the rock-paper-scissors model, where species dominance follows the rules: scissors cut paper, paper wraps rock, and rock crushes scissors. The three species are denoted by $i$, with $i=1,2,3$, as illustrated in Fig.~\ref{fig1}, where blue, pink, and purple stand for species $1$, $2$, and $3$, respectively. As indicated by the arrows, organisms of species $i$ eliminate individuals of species $i+1$, with $i=i\,+\,3\,\alpha$, where $\alpha$ is an integer. In our study, a contagious disease spreads through the system, transmissible person-to-person. All organisms of every species are equally susceptible to virus infection. If contaminated, individuals may be cured or die because of disease complications; in case of becoming healthy again, there is no immunity, meaning reinfection is possible at any time.

Our model considers that organisms of one out of the species evolve to 
perform a self-preservation strategy when perceiving the arrival of a disease outbreak in their neighbourhood. The behavioural method restricts mobility whenever sick individuals' density exceeds a tolerable threshold. Performing the locally adaptive mobility limitation consists of: i) scanning the environment;
ii) identifying the neighbours which are viral vectors; iii)
calculating the density of infected individuals in the organisms' neighbourhood; iv) decelerating if the local density of ill individuals is higher than a tolerable threshold. This means that each organism is independent in reducing its velocity when avoiding infection or moving with maximum speed to explore the territory.

\subsection{Simulations}

We use stochastic simulations of the rock-paper-scissors model, using the May-Leonard implementation, where the total number of organisms is not conserved \cite{leonard}. The lattices created to run the simulations are square with periodic boundary conditions, which means that organisms interact on a torus surface with $\mathcal{N}$ points. At most, one individual can occupy each grid site; thus, the maximum number of individuals is $\mathcal{N}$. 

The density of individuals of species $i$, $\rho_i(t)$, with $i=1,2,3$, is the fraction of the lattice occupied by individuals of the species $i$ at time $t$:
\begin{equation}
\rho_i(t)=\frac{I_i(t)}{\mathcal{N}},
\end{equation}
where $I_i(t)$ is the total number of organisms of species $i$ at time $t$. Therefore, the density of empty spaces is given by $\rho_0 = 1 - \rho_1 - \rho_2 - \rho_3$.

Initially, each individual is allocated at a random grid point, with the initial density of individuals being the same 
for every species: $\rho_i \approx \mathcal{N}/3$, with $i=1,2,3$. Throughout this work, all simulations are performed assuming the initial number of individuals as the maximum integer number that fits on the lattice, $I_i (t=0)\,\approx \,\mathcal{N}/3$, with $i=1,2,...,3$. The remaining grid sites are left empty in the initial state.
Furthermore, the initial proportion of sick individuals is $1\%$, which is valid for every species.

We use the notation $h_i$ and $s_i$ to identify healthy and sick individuals of species $i$; the labelling $i$ stands for all individuals, irrespective of illness or health. The stochastic interactions can be illustrated as follows:
\begin{itemize}
\item 
Selection: $ i\ j \to i\ \otimes\,$, with $ j = i+1$, where $\otimes$ means an empty space. The grid site previously occupied by the individual of species $i+1$ becomes empty due to the selection interaction.
\item
Reproduction: $ i\ \otimes \to i\ i\,$. An offspring of any species can occupy available empty space.
\item 
Mobility: $ i\ \odot \to \odot\ i\,$, where $\odot$ means either an individual of any species or an empty site. An organism of species $i$ switches position with another individual of any species or with an empty space.
\item 
Infection: $ s_i\ h_j \to s_i\ s_j\,$, with $i,j=1,2,3$. An ill individual of species $i$ transmits the disease to a healthy individual of any species.
\item 
Cure: $ s_i \to h_i\,$. An ill organism of species $i$ is cured of the disease, remaining susceptible to reinfection.
\item 
Death: $ s_i \to \otimes\,$. A sick individual of species $i$ dies, 
leaving its position empty.
\end{itemize}
The probability of a given interaction occurring depends
on the set of real parameters: $S$ (selection rate), $R$ (reproduction rate), $M$ (mobility rate), $\kappa$ (infection rate), $\mu$ (mortality rate), and $C$ (cure rate). 

\subsection {Implementation of the Strategic Mobility Limitation}

Because of the behavioural strategy of limiting movement rhythm to minimise the chances of contamination, the mobility of individuals of species $1$ is implemented as follows:
\begin{enumerate}
\item
it is introduced the organisms' perception radius, $\mathcal{R}$, which indicates the maximum distance an organism of species $1$ can observe the environment to be aware of the arrival of a local outbreak;
\item
it is computed the local density of sick organisms of any species
within a circular area of radius $\mathcal{R}$, centred in the organism of species $1$;
\item
it is defined the mobility restriction trigger, $\varphi$, a real parameter in the interval $0\leq \varphi \leq 1$ to characterise the minimum local density of sick organisms that imposes mobility limitation;
\item
it is defined a slowness factor $\nu$, with $0\leq \nu \leq 1$,
a real parameter defining the percentage reduction in the mobility rate;
\item
if the local density of sick organisms is lower than $\varphi$, the individual walks with probability is $m$; otherwise the effective mobility rate is $(1-\nu)\,m$.
\end{enumerate}

We work with the Moore neighbourhood, with an organism interacting with one of its eight immediate neighbours. The numerical code follows the steps: i) randomly choosing an active individual among all individuals in the lattice; ii) drawing one interaction to be executed; iii) raffling one of the eight immediate neighbours to suffer the interaction. 
One time step is computed if either an interaction is implemented or an organism of species $1$ purposely remains static due to the mobility restriction strategy. After $\mathcal{N}$ occurs, one generation, our time unit, is completed.

Throughout this paper, all outcomes were obtained from running simulations in lattices with $500^2$ sites, for a timespan of $5000$ generations. We set the parameters to $S = R=1.0$, $M=2.0$, and $C=0.2$; the perception radius is assumed to be $\mathcal{R}=3$. The disease infection and mortality rates are chosen according to the experiments described in the following sections. However, we ran thousands of simulations for different parameters, confirming that our conclusions are valid for other interaction rates and perception radii.

\section{Organisms' spatial organisation}
\label{sec3}
Let us study how the mobility restriction of individuals of species $1$ affects the spatial patterns arising from random initial conditions. For this purpose, we first performed three simulations with different disease virulences: i) Simulation A:
$\kappa=2.0$ and $\mu=0.2$; ii) Simulation B:
$\kappa=2.0$ and $\mu=0.8$;  iii) Simulation C:
$\kappa=4.0$ and $\mu=0.2$. The mobility restriction trigger is $\varphi=0.1$ and the slowness factor is $\nu=0.9$.

\begin{figure}[h]
\centering
       \begin{subfigure}{.48\textwidth}
        \centering
        \includegraphics[width=85mm]{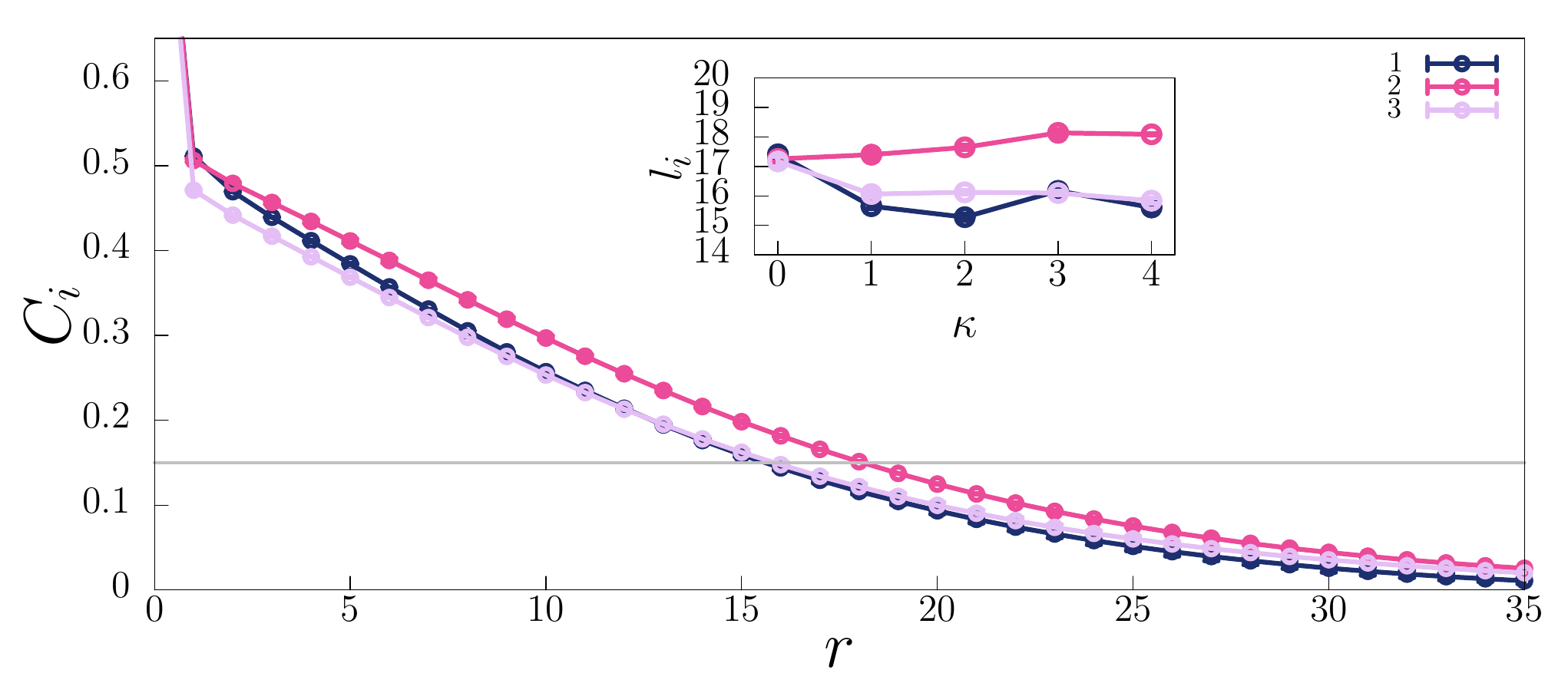}
        \caption{}\label{fig6a}
    \end{subfigure}\\
           \begin{subfigure}{.48\textwidth}
        \centering
        \includegraphics[width=85mm]{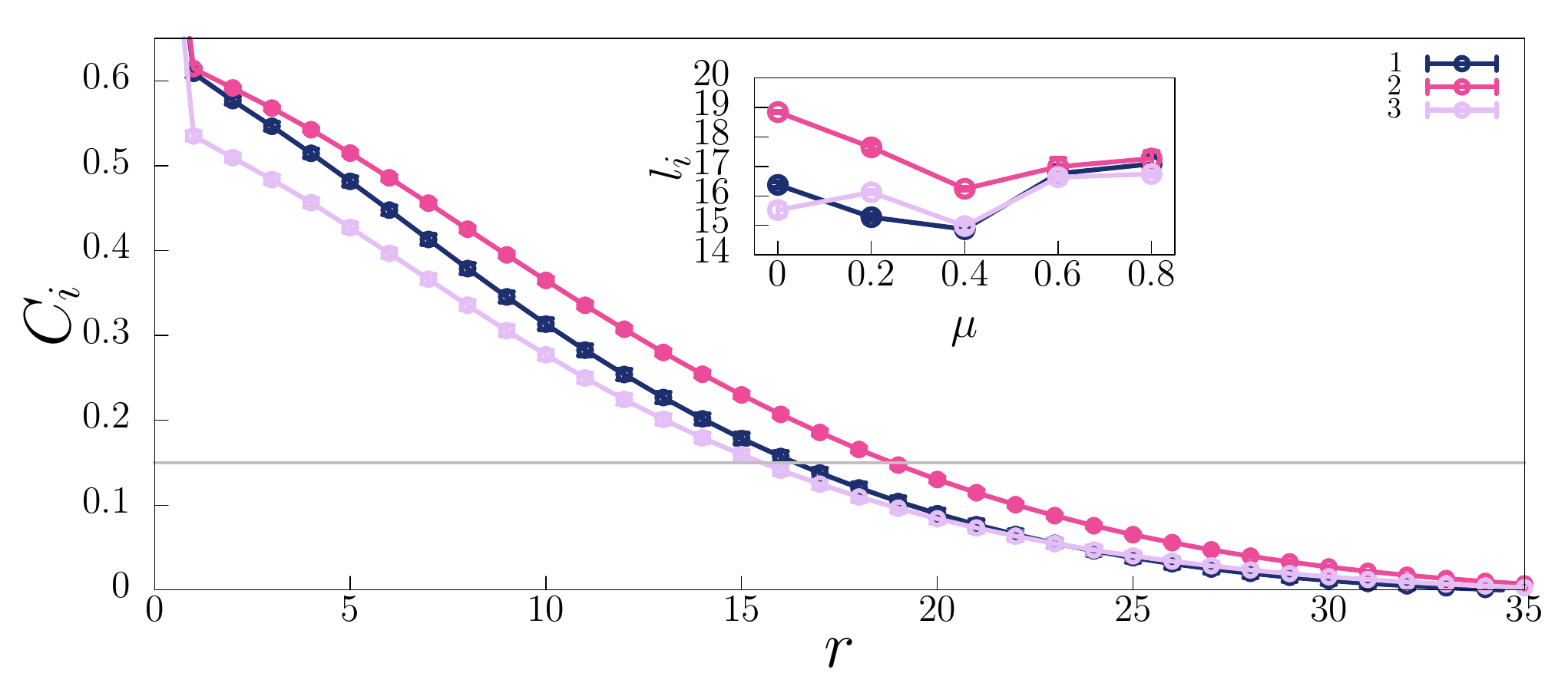}
        \caption{}\label{fig6b}
    \end{subfigure}
\caption{Autocorrelation function in the rock-paper-scissors model with mobility restriction tactic of organisms of species $1$. The outcomes were obtained by averaging sets of $100$ simulations for $\varphi=0.1$ and $\nu=0.9$; the error bars show the standard deviation. Figure \ref{fig6a} depicts $C_i$ for $\kappa=4.0$ and $\nu=0.2$, while Fig. \ref{fig6b} shows the outcomes for $\kappa=2.0$ and $\nu=0.1$. Blue, pink, and purple lines represent species $1$, $2$, and $3$, respectively. The inset panels in Figs.~\ref{fig6a} and \ref{fig6b} depict the characteristic length scale of the typical spatial domain of each species as a function of the disease infection and mortality rates, respectively.}
\label{fig6}
\end{figure}

The final individuals' spatial organisation is shown in Figs. \ref{fig2a} (Simulation A), \ref{fig2c} (Simulation B), and \ref{fig2e} (Simulation C), respectively. Dots depict the individuals according to the scheme in Fig.~\ref{fig1}: blue, pink, and purple indicate individuals of species $1$, $2$, and $3$, respectively. Additionally, the spatial distribution of the density of sick organisms in the final snapshot of Simulations A, B, and C are shown in Figs. \ref{fig2b}, Fig. \ref{fig2d}, and Fig. \ref{fig2f}, respectively. The colour bar shows the density of sick organisms at each grid site. 
See also the dynamics of the spatial patterns during Simulations A, B, and C in videos https://youtu.be/mSsL0XfMtlI, https://youtu.be/Iwy25cR9bsM, and https://youtu.be/2a1pieelQrc; the respective local outbreaks appear in videos https://youtu.be/i7q9RsI0ZW0,
https://youtu.be/T1vnJ2aRBgQ, https://youtu.be/ll17ZnGh1cc.

Finally, the temporal variation of the density of individuals of species $i$ and empty spaces during Simulations A, B, and C are depicted in Figs.~\ref{fig3a}, ~\ref{fig3b}, and ~\ref{fig3c}. Blue, pink, and purple lines depict the fraction of the grid with organisms of species $1$, species $2$, and species $3$, respectively; the orange line shows the density of empty spaces. 
The outcomes show that individuals of the same species segregate in departed spatial domains after an initial pattern formation stage. The arising of such single-species regions reduces the density of empty spaces, which grows initially because of the random organisms' distribution.

Figures \ref{fig2a} and \ref{fig2b} show that the disease virulence is sufficient to produce intense disease outbreaks in 
Simulation A. Therefore, the self-protective strategy performed by individuals of species $1$ is commonly triggered, leading to the formation of asymmetric spiral patterns. 
This happens because, on average, individuals of species $1$ explore the smallest area of the lattice per unit time, which is proportional to the effective mobility rate according to the 
random walk theory \cite{mobilia2,random}. This unbalances the rock-paper-scissors model, allowing species $2$ to dominate, as shown in Fig. \ref{fig3c} \cite{unevenmobility}.

Figures \ref{fig2c} and \ref{fig2d} reveal that if a virus mutation makes the disease more mortal ($\mu=0.2$ changes to $\mu=0.8$), the number of sick organisms significantly decreases. Thus, the mobility restriction strategy is performed by fewer organisms. Because of this, the asymmetry verified in the spirals in Simulation B is reduced compared with Simulation A. Consequently, the difference among the average species densities is less relevant, as shown in Fig. \ref{fig3b}. 
In contrast, according to Figs. \ref{fig2e} and \ref{fig2f}, if the disease becomes more contagious, the ($\kappa=1.0$ changes to $\kappa=4.0$), the average density of sick organisms rises. This means that the vast majority of the organisms of species $1$ remain static to avoid infection, which accentuates the asymmetry in spiral patterns and, consequently, in species populations, as depicted in Fig. \ref{fig3c}. 

To study how mobility restrictions decelerate epidemic spreading among individuals in more detail, we performed another simulation starting from the prepared spatial configuration in Fig.~\ref{fig4a}. Accordingly, organisms of a single species occupy torus ring surfaces with the same width: from left to right, one has species $3$ (purple), $1$ (blue), and $2$ (pink). The fraction of sick individuals randomly allocated within each group is $1\%$, which results in the initial local density of sick organisms shown in Fig.~\ref{fig4f}. We refer to this realisation as Simulation D, implemented using the parameters $\mu=0.2$, $\kappa=4.0$, $\varphi=0.1$, and $\nu=0.9$. The dynamics of organisms' spatial organisation and local epidemic surges are shown in videos https://youtu.be/WTa-QUa3NFw and https://youtu.be/VZkpNzW2-cs.

Firstly, we observe the initial stage of Simulation D shown in \ref{fig4b} to \ref{fig4e} for $t=20$, $t=40$, $t=60$, and $t=100$; the respective spatial distribution of the local disease surges appears in Figs. \ref{fig4g} to \ref{fig4j}. As the virus spreads among individuals of the same species, the number of empty spaces grows because of deaths caused by the disease. However, epidemics spread slower among organisms of species $1$ because of the movement limitation response. One sees in Figs. \ref{fig4h}, \ref{fig4i}, and \ref{fig4j} that the density of ill individuals is lower in the centre of the lattice.

Secondly, let us observe the final stage of Simulation D.
Individuals and local density of sick organisms at $t=2000$, $t=3000$, $t=3140$, $t=3400$, and $t=3920$ are shown in Figs. \ref{fig5a} to \ref{fig5e}, and \ref{fig5f} to \ref{fig5j}, respectively. As Simulation D proceeds, organisms of species $i$ kill organisms of species $i+1$. This results in the torus surface rotation from left to right, as shown in Figs.~\ref{fig5a} and ~\ref{fig5f}. At this point, many sick individuals are eliminated on the borders of the spatial domain, and new healthy organisms appear (all offspring are healthy). Therefore, the average density of sick individuals decreases when compared with the initial stage of the simulation.

As the average mobility of individuals of species $1$ is lower than other species, we observe a narrowing of the blue ring and, consequently, the purple one \cite{unevenmobility}. As shown in Figs.~\ref{fig5b} and \ref{fig5g}, organisms of species $1$ manage to move without being caught by individuals of species $3$, reaching the area dominated by species $2$. From this point on, species $1$ proliferates (blue) by killing individuals of species $2$ (pink), producing a spiral wave that spreads on the lattice, as appears in Figs. \ref{fig5c} to \ref{fig5e}. 
As the wave propagates, selection interactions result in the death of many infected organisms, which are further substituted by healthy offspring. Therefore, the average density of sick organisms drops, lowering the intensity of local epidemic surges - Figs. \ref{fig5h} to \ref{fig5j}.

\section{Typical spatial domain's characteristic length scales}
\label{sec4}
The mobility restriction tactic to local disease outbreaks of organisms of species $1$ introduces an asymmetry in the spiral patterns, as observed in 
the previous section. We now aim to quantify the average characteristic length of the typical spatial domains occupied by individuals of each species. 
For this purpose, we calculate the spatial autocorrelation function $C_i(r)$, with $i=1,2,3$, in terms of radial coordinate $r$ by introducing the function $\phi_i(\vec{r})$ that identifies the positions in the lattice occupied by individuals of species $i$ (healthy and sick individuals). Using the Fourier transform
\begin{equation}
\Phi_i(\vec{K}) = \mathcal{F}\,\{\phi_i(\vec{r})-\langle\phi_i\rangle\},
\end{equation}
where $\langle\phi_i\rangle$ is the mean value of $\phi_i(\vec{r})$, we find the spectral densities
\begin{equation}
S_i(\vec{K}) = \sum_{K_x, K_y}\,\Phi_i(\vec{K}).
\end{equation}

Therefore,
\begin{equation}
C_i(\vec{r}') = \frac{\mathcal{F}^{-1}\{S_i(\vec{K})\}}{C(0)},
\end{equation}
which is written in terms of the radial coordinate $r$ as
\begin{equation}
C_i(r) = \sum_{|\vec{r}'|=x+y} \frac{C_i(\vec{r}')}{min\left[2N-(x+y+1), (x+y+1)\right]}.
\end{equation}

Our statistics are produced using sets of $100$ stochastic simulations with different initial conditions for $\varphi=0.1$ and $\nu=0.9$. We assume the threshold $C(l)=0.15$ to determine the characteristic length scale, $l_i$, of the typical spatial domains of species $i$.
Figures \ref{fig6a} and \ref{fig6b} depict the average autocorrelation function for the disease virulence parameters ($\kappa=4.0$; $\mu=0.2$) and ($\kappa=2.0$; $\mu=0.1$), respectively. In both scenarios, highly contagious disease outbreaks spread through the lattice. The blue, pink, and purple lines depict the autocorrelation function of individuals of species $1$, $2$, and $3$, respectively. The error bars indicate the standard deviation; the horizontal grey line represents the threshold used to calculate the characteristic length scale.

Due to the locally adaptive movement restriction response executed by individuals of species $1$, organisms of species $2$ are more spatially correlated. The inset figure in Fig. \ref{fig6a} depicts the characteristic length for several disease transmission rates for $\mu=0.2$, namely, $1.0 \leq \kappa \leq 4.0$, in intervals of $\Delta \kappa =1.0$. 
In addition, $l_i$ for a range of mortality rates for $\kappa=2.0$ appears in the inset Fig. \ref{fig6b}, with $0.0 \leq \mu \leq 0.8$, in intervals of $\Delta \mu =0.2$. 
The outcomes show that the mobility limitation in $90\%$ assumed by organisms of species $1$ to minimise infection risk whenever the local density of sick individuals exceeds $10\%$ results in an asymmetric formation of groups of conspecifics. Namely, species $2$ occupies the most extensive areas, with unevenness accentuating as
either the disease is transmitted more rapidly ($\kappa$ grows) or causes fewer deaths ($\nu$ decreases).
Because of reduced mobility, organisms of species $1$ live in the smallest spatial domains, except in the case of nonlethal disease ($\mu=0.0$), where species $3$ are constrained to populate areas with the shortest characteristic length scale.

\section{Impact of disease virulence at individual and population levels}
\label{sec5}

\begin{figure}
 \centering
       \begin{subfigure}{.48\textwidth}
        \centering
        \includegraphics[width=85mm]{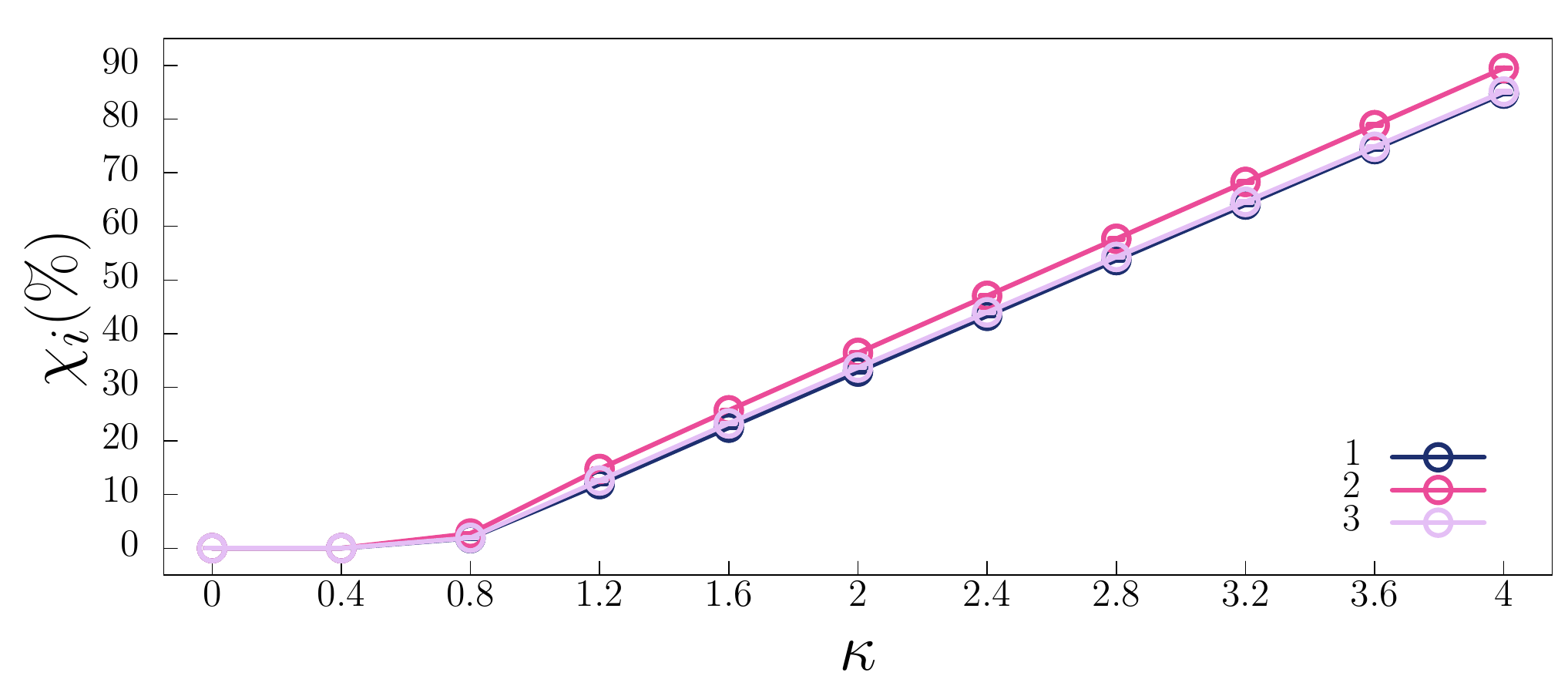}
        \caption{}\label{fig7a}
    \end{subfigure}\\
           \begin{subfigure}{.48\textwidth}
        \centering
        \includegraphics[width=85mm]{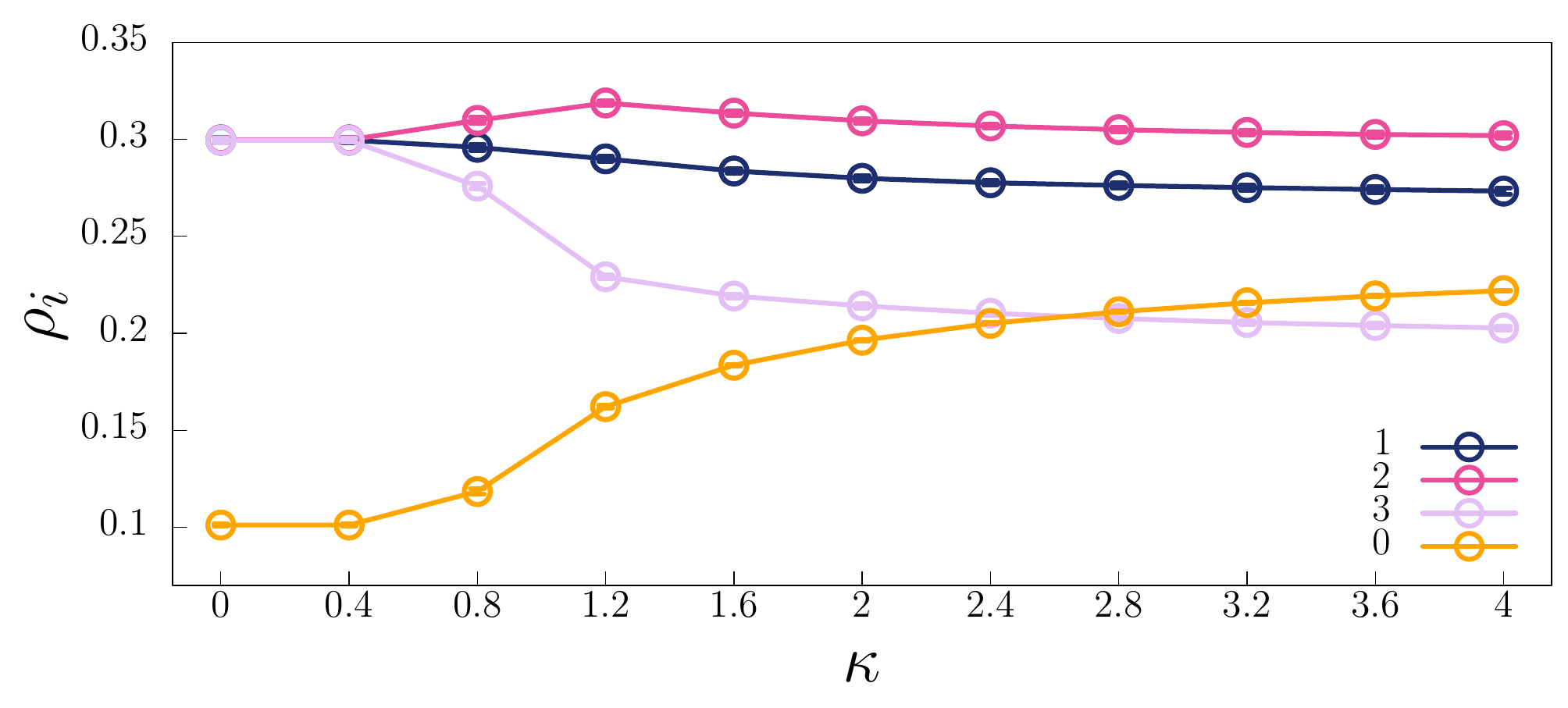}
        \caption{}\label{fig7b}
    \end{subfigure}
\caption{Organisms' infection risk and species densities in terms of the disease transmission rate.
Figure \ref{fig7a} shows $\chi_i$, where blue, pink, and purple lines stand for species $1$, $2$, and $3$, respectively. Figure \ref{fig7b} depict $\rho_i$, for species $1$, $2$, and $3$, and the density of empty spaces $\rho_0$ (orange line).
The results were averaged from sets of $100$ simulations with different random initial conditions; the error bars indicate the standard deviation. The parameters are $\varphi=0.1$ and $\nu=0.9$.}
  \label{fig7}
\end{figure}

We now simulate a series of simulations for a
range of infection and mortality rates to quantify the impact of varying disease virulence at population and individual levels. 
For this purpose, we calculate the species densities $\rho_i$ and the average organisms' infection risk $\chi_i(t)$, defined as the probability of a healthy organism of species $i$ being contaminated at time $t$. 
To implement $\chi_i$, the code follows the steps:
i) counting the total number of healthy individuals of species $i$ when each generation begins; ii) computing the number of healthy individuals of species $i$ infected during the generation; 
iii) calculating the infection risk, $\chi_i$, with $i=1,2,3$, defined as the ratio between the number of infected individuals and the initial number of healthy individuals of species $i$.

We ran two sets of simulations: 
\begin{itemize}
\item
Varying transmission rate: $0 \leq \kappa \leq 4.0$, in intervals of $\Delta \kappa =0.4$, for fixed mortality rate $\mu=0.2$; \item
Varying mortality rate: $0 \leq \mu \leq 0.8$, in intervals of $\Delta \mu =0.08$, for fixed infection rate $\kappa=2.0$. 
\end{itemize}
We conducted the statistical analyses by avoiding high fluctuations inherent to the pattern formation process by computing the mean infection risk and species densities using the data from the second simulation half. We assume $\varphi=0.1$ and $\nu=0.9$.

\begin{figure}
 \centering
       \begin{subfigure}{.48\textwidth}
        \centering
        \includegraphics[width=85mm]{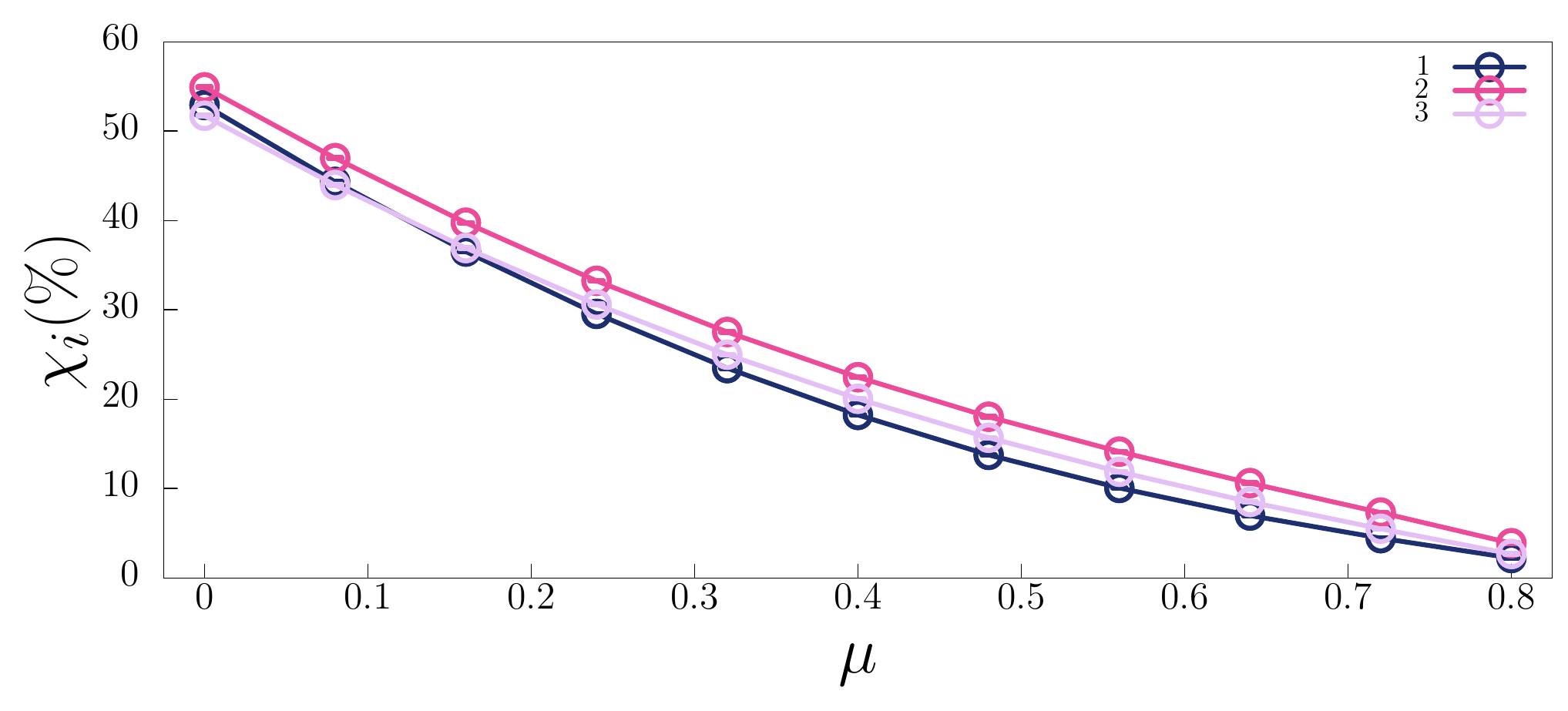}
        \caption{}\label{fig8a}
    \end{subfigure}\\
           \begin{subfigure}{.48\textwidth}
        \centering
        \includegraphics[width=85mm]{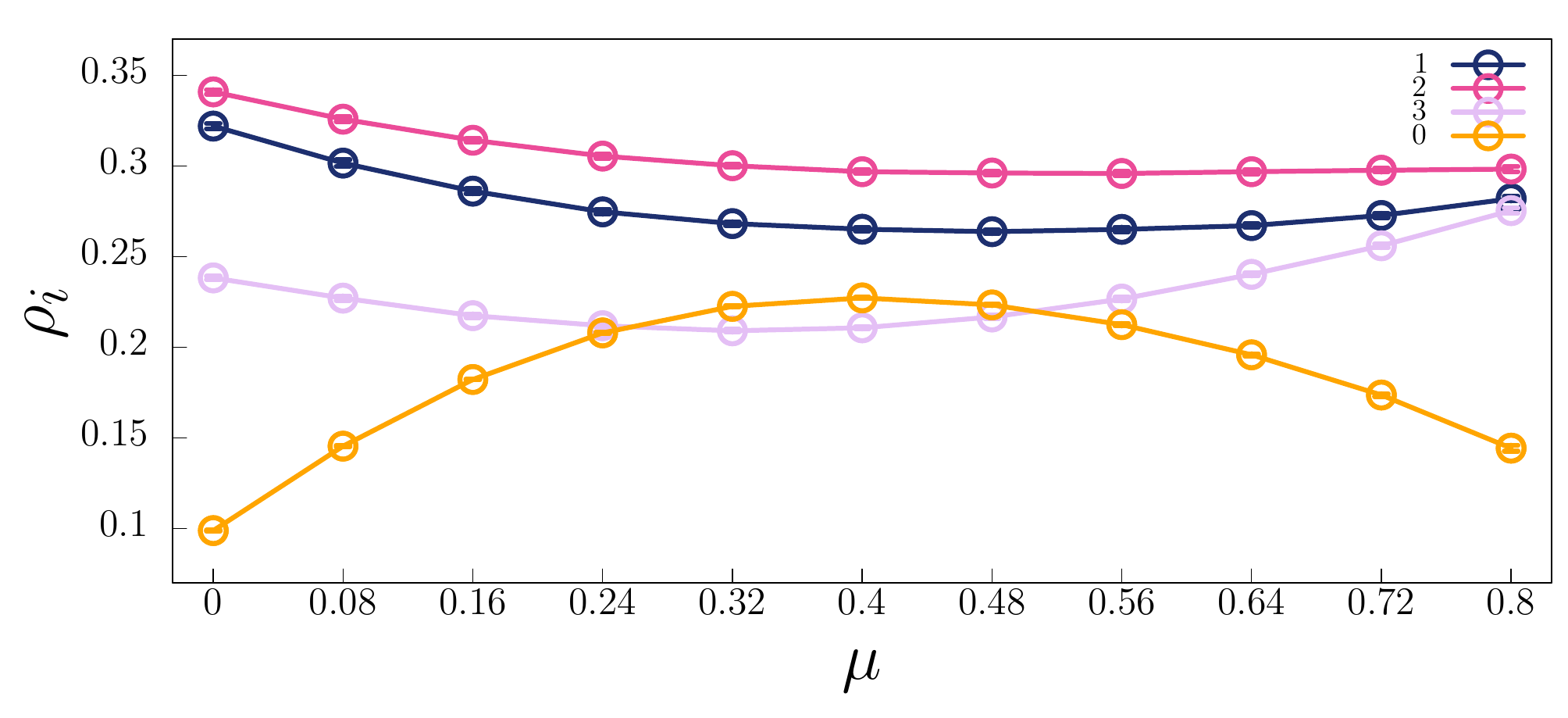}
        \caption{}\label{fig8b}
    \end{subfigure}
\caption{Organisms' infection risk and species densities as function of the disease mortality rate.
Figure \ref{fig8a} and \ref{fig8b} depict $\chi_i$ and $\rho_i$ , with blue, pink, and purple lines indicating the species $1$, $2$, and $3$, respectively; the orange line in Fig~ \ref{fig8b} shows the variation of the density of empty spaces with $\mu$.
The error bars indicate the standard deviation from sets of $100$ simulations; we used $\varphi=0.1$ and $\nu=0.9$.}
  \label{fig8}
\end{figure}

Figures \ref{fig7a} and \ref{fig7b} show the dependence of the organisms' infection risk and species densities 
in the transmission rate $\kappa$, while Figs. \ref{fig8a} and \ref{fig8b} depict the outcomes for a varying mortality rate. The colours follow the scheme in Fig.~\ref{fig1}: blue, pink, and purple lines depict the results for species $1$, $2$, and $3$, respectively. The orange lines in Figs. \ref{fig7b} and \ref{fig8b} represents the density of empty spaces; the error bars indicate the standard deviation.

The organisms' infection risk rises if pathogen mutations cause a more contagious or less lethal disease. This happens because the local density of sick organisms increases as $\kappa$ grows and $\mu$ drops, as observed in Fig.~\ref{fig2}. Therefore, mobility limitation in $90\%$ of individuals of species $1$ is more triggered as $\kappa$ increases and $\mu$ decreases, which accentuates unevenness in the spatial rock-paper-scissors game. Furthermore, the outcomes reveal that for non-lethal diseases ($\mu=0.0$), organisms of species $1$ are no longer less likely to be contaminated even properly reacting to local surges; instead, the 
infection risk for individuals of the species $2$ is the lowest.

For $\kappa \leq 0.4$, the disease transmission is insufficient to generate a disease surge spreading on the lattice, given that the initial fraction of ill individuals is $1\%$. For $\kappa \geq 0.8$,
the infection risk increases, with organisms of species $1$ being less likely to be contaminated due to the mobility restriction strategy. Being less mobile, the chances of an organism of species $1$ finding and eliminating individuals of species $2$ reduces; this leads species $2$ to predominate in the spatial game. According to the results, the density of empty spaces reaches the maximum value for $\mu = 0.4$, which is the mortality rate producing the most relevant decline in the species populations. Nevertheless, if the disease mortality becomes high, the virus spreads slower since most ill individuals die before passing the pathogen to healthy neighbours.

\begin{figure}
 \centering
       \begin{subfigure}{.48\textwidth}
        \centering
        \includegraphics[width=85mm]{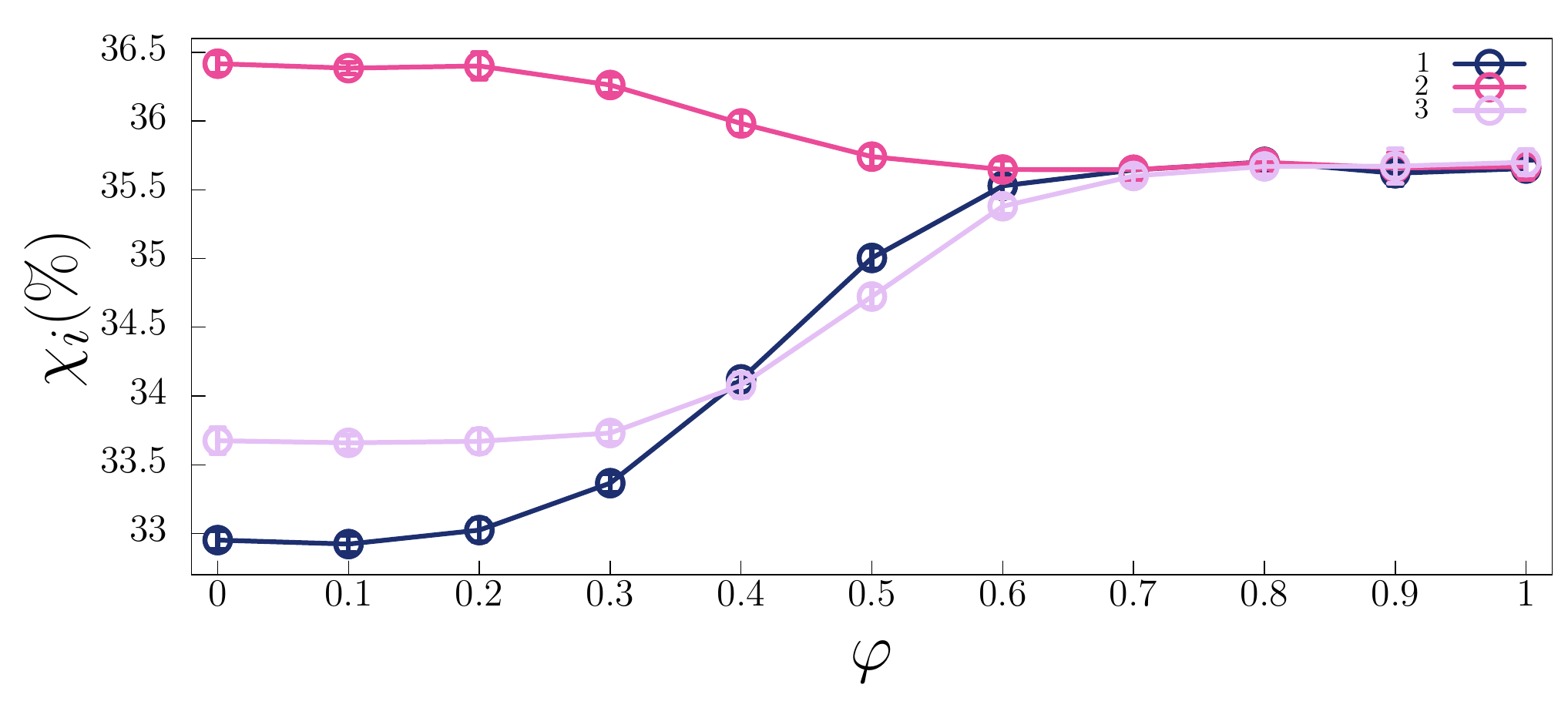}
        \caption{}\label{figAa}
    \end{subfigure}\\
           \begin{subfigure}{.48\textwidth}
        \centering
        \includegraphics[width=85mm]{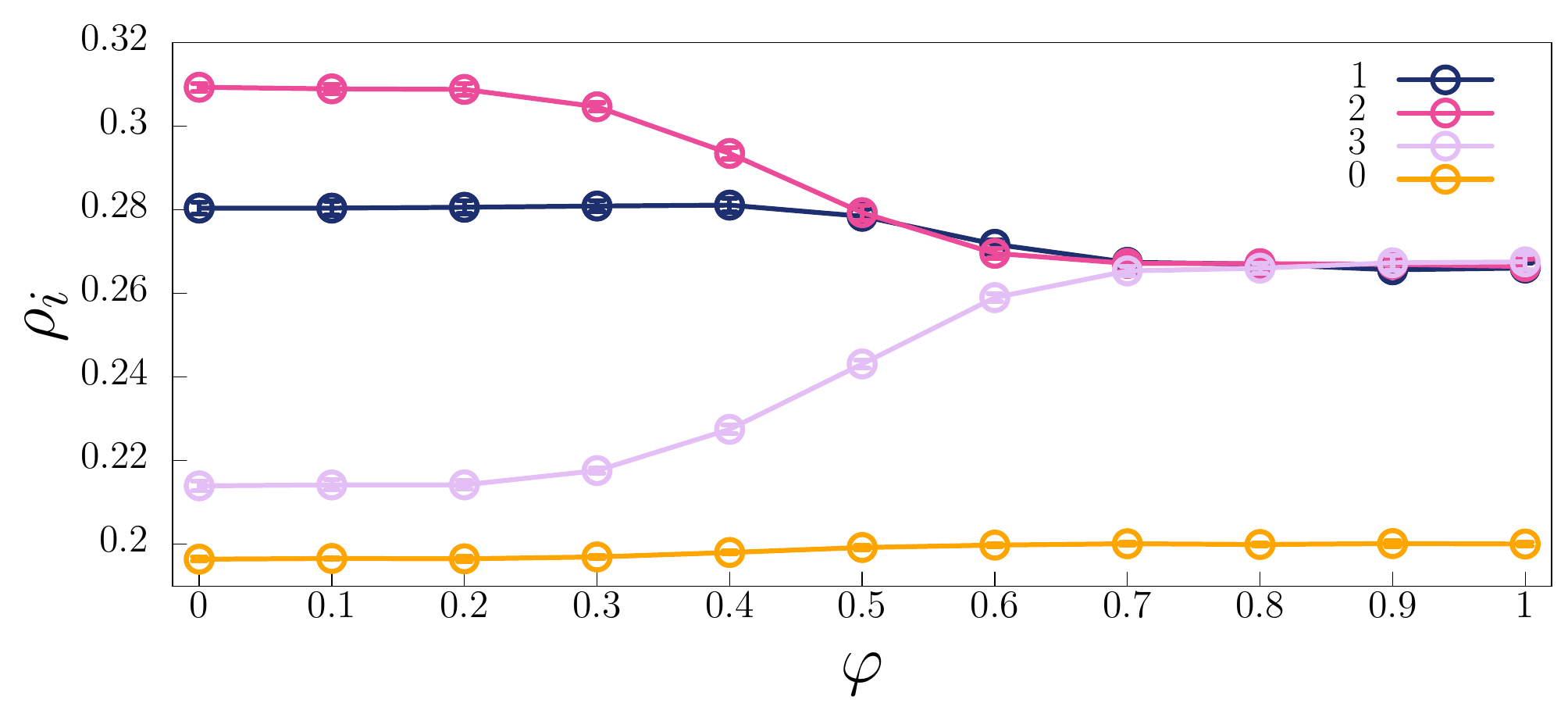}
        \caption{}\label{figAb}
    \end{subfigure}
\caption{Organisms' infection risk and species densities in terms of the disease transmission rate.
Figure \ref{fig7a} shows $\chi_i$, where blue, pink, and purple lines stand for species $1$, $2$, and $3$, respectively. Figure \ref{fig7b} depict $\rho_i$, for species $1$, $2$, and $3$, and the density of empty spaces $\rho_0$ (orange line).
The results were averaged from sets of $100$ simulations with different random initial conditions; the error bars indicate the standard deviation. The parameters are $\varphi=0.1$ and $\nu=0.9$.}
  \label{figA}
\end{figure}

\section{Role of mobility restriction trigger and slowness factor}
\label{sec6}

We now investigate the influence of the mobility restriction trigger and the slowness factor at individual and population levels. 
Considering the transmission and mortality rates are fixed, namely, $\kappa=2.0$ and $\mu=0.2$, we quantified the effects of the mobility unevenness controlled by $\varphi$ and $\nu$ in the selection risk and species densities.

\begin{figure}
 \centering
       \begin{subfigure}{.48\textwidth}
        \centering
        \includegraphics[width=85mm]{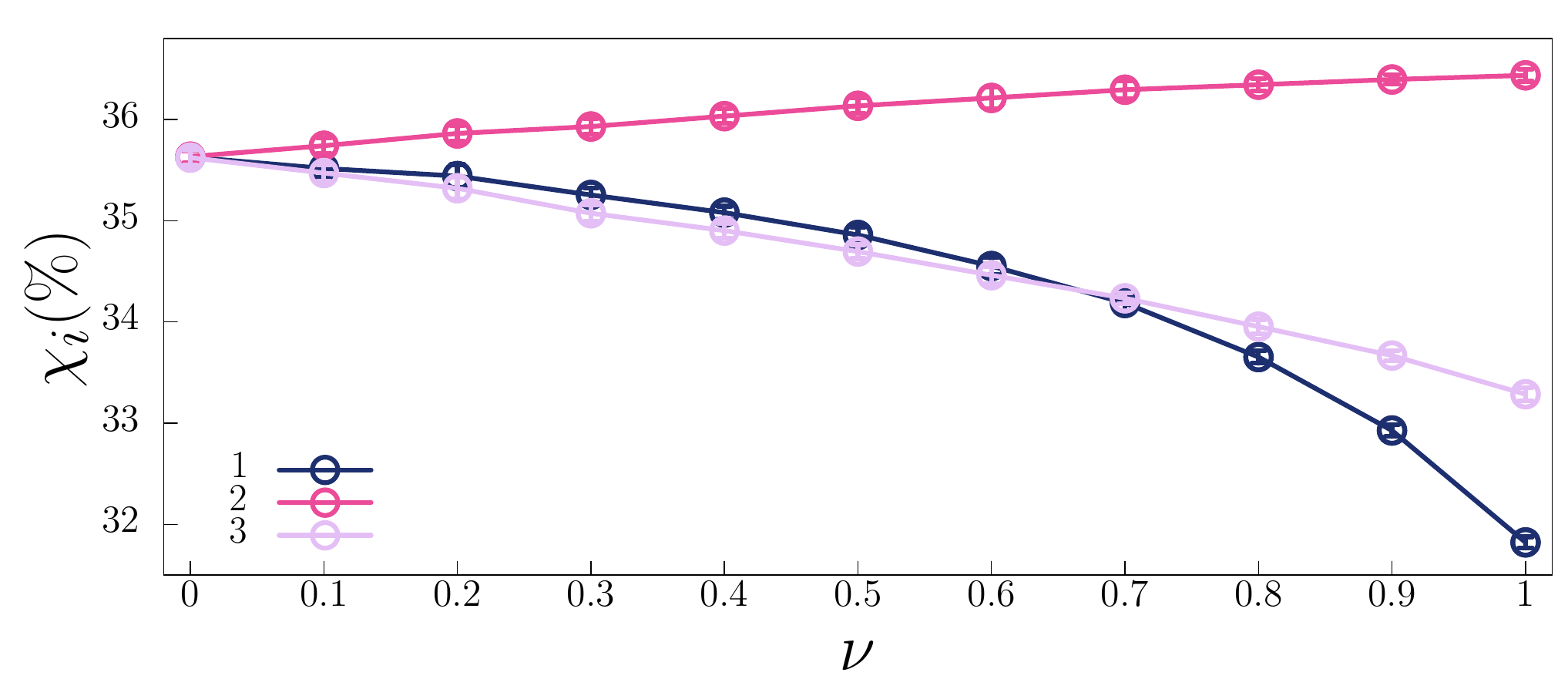}
        \caption{}\label{figBa}
    \end{subfigure}\\
           \begin{subfigure}{.48\textwidth}
        \centering
        \includegraphics[width=85mm]{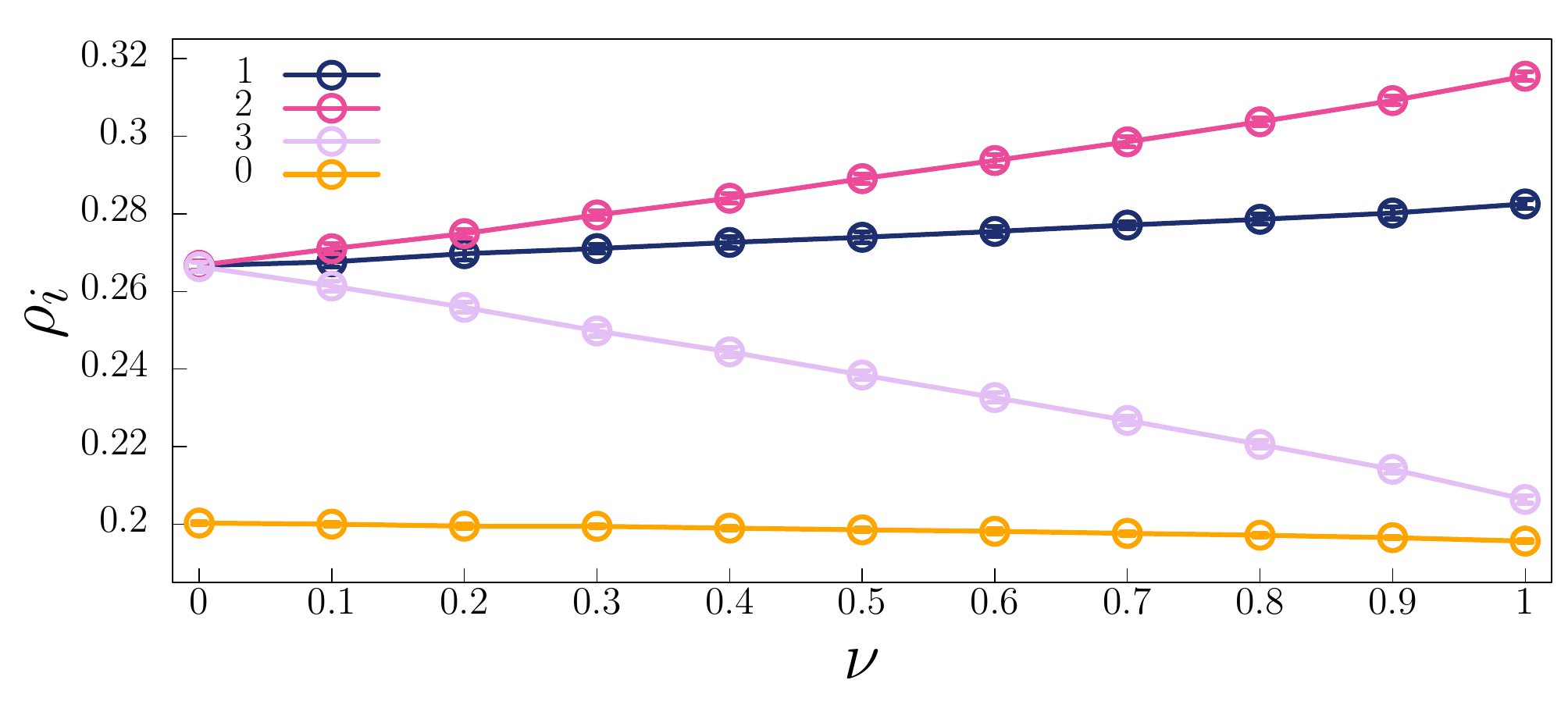}
        \caption{}\label{figBb}
    \end{subfigure}
\caption{Organisms' infection risk and species densities in terms of the disease transmission rate.
Figure \ref{fig7a} shows $\chi_i$, where blue, pink, and purple lines stand for species $1$, $2$, and $3$, respectively. Figure \ref{fig7b} depict $\rho_i$, for species $1$, $2$, and $3$, and the density of empty spaces $\rho_0$ (orange line).
The results were averaged from sets of $100$ simulations with different random initial conditions; the error bars indicate the standard deviation. The parameters are $\varphi=0.1$ and $\nu=0.9$.}
  \label{figB}
\end{figure}

The infection risk and species density in terms of the mobility restriction trigger are depicted in Figs. \ref{figAa} and \ref{figAb}, while the dependence of $\chi_i$ and $\rho_i$ on the slowness factor appear in Figs. \ref{figBa} and \ref{figBb}.
Blue, pink, and purple lines depict the results for species $1$, $2$, and $3$, respectively; the orange lines indicate the density of empty spaces. The outcomes were obtained by performing sets of $100$ simulations; the error bars indicate the standard deviation.

Figure \ref{figA} reveal that if organisms of species $1$ decelerate only if the local density of sick organisms exceeds $70\%$, the behavioural tactic does not induce significant unevenness in the spatial game. As the mobility restriction trigger is lowered, the chances of organisms of species $1$ and $2$ being contaminated decrease, with $\chi_1$ being the lowest for $\varphi <0.4$. Furthermore, the outcomes reveal that the species densities remain approximately constant for $\varphi \leq 2$.

Figure \ref{figB} shows that, although $\chi_1$ decreases even for a slight deceleration ($\nu \gtrsim 0$), organisms of species $1$ are the ones with the lowest risk of being contaminated if $\nu \geq 0.7$ - for $\nu < 0.7$, individuals of species $3$ are the less likely to become sick. 
Concerning the cyclic competition for space, 
the slower organisms of species $1$ become when perceiving a local epidemic outbreak, the more accentuated the mobility unevenness introduced in the spatial rock-paper-scissors game. Thus, the lower the slowness factor, the more the prevalence of species $2$ becomes significant.

\section{Effectiveness of the mobility restriction strategy}
\label{sec7}

We now study the influence of the locally adaptive mobility restriction control parameters - $\varphi$ and $\nu$ - on the effectiveness of the organisms' behavioural strategy. In this study, we compute the relative decrease in individuals' contamination risk for a range of mobility restriction trigger and slowness factors and $\nu$ for various disease virulence scenarios.

\begin{figure}[t]
 \centering
       \begin{subfigure}{.48\textwidth}
        \centering
        \includegraphics[width=85mm]{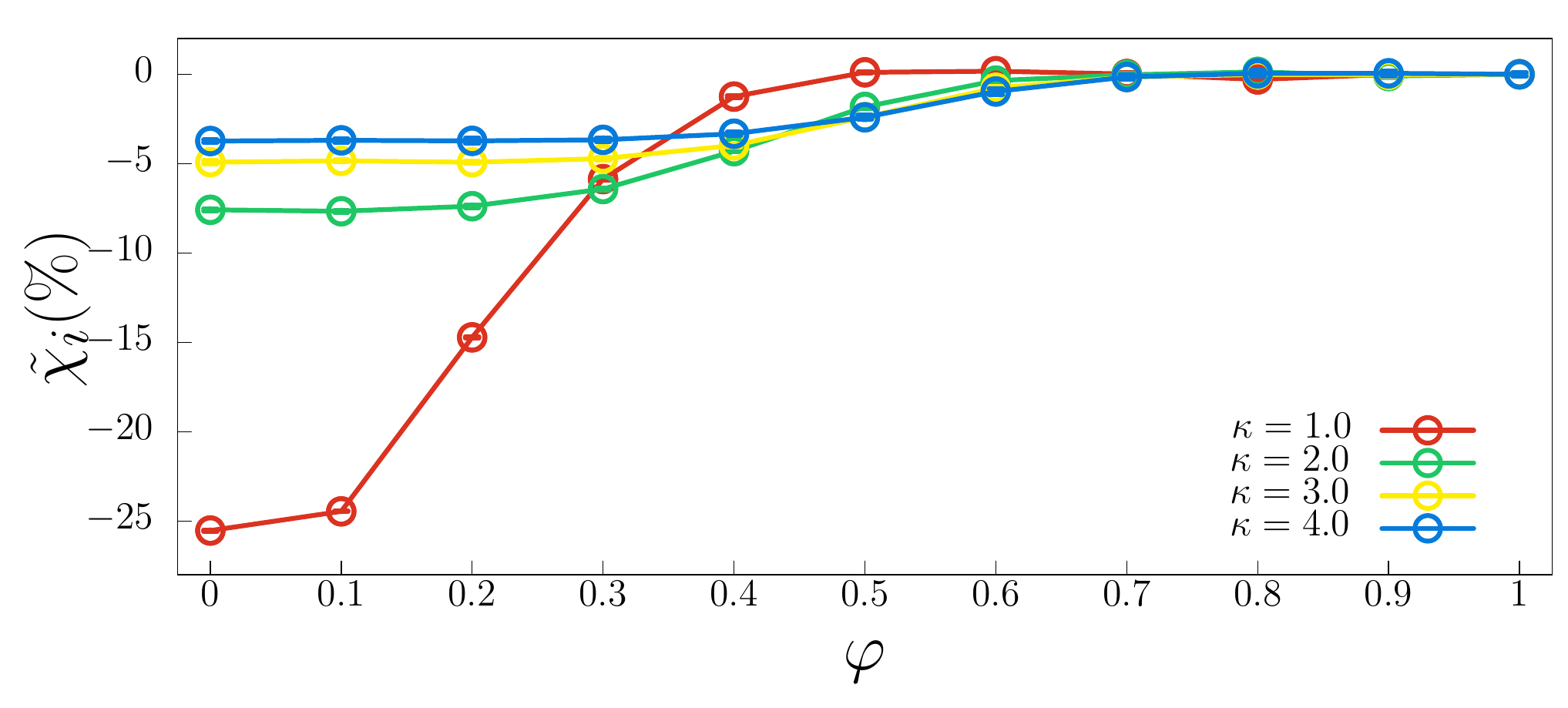}
        \caption{}\label{fig9a}
    \end{subfigure}\\
           \begin{subfigure}{.48\textwidth}
        \centering
        \includegraphics[width=85mm]{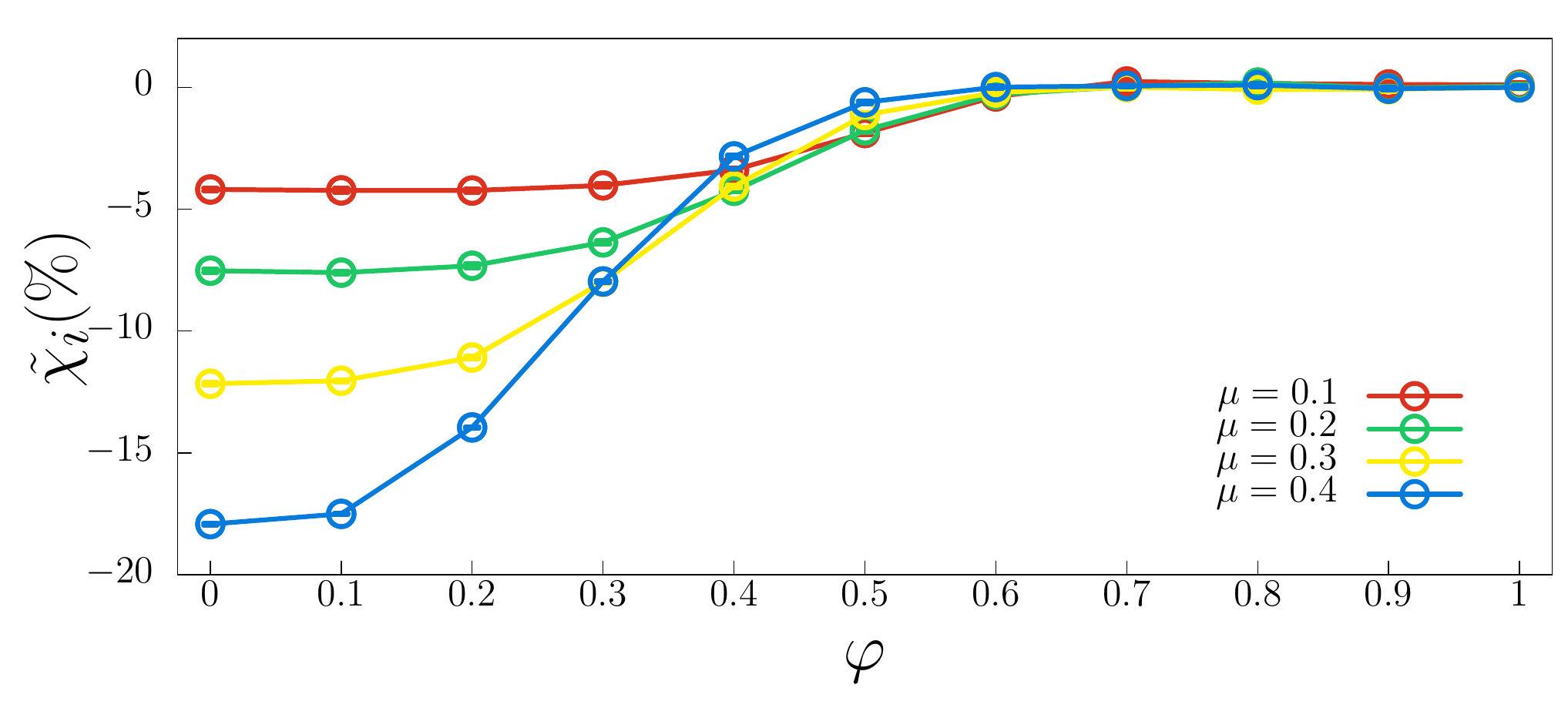}
        \caption{}\label{fig9b}
    \end{subfigure}
\caption{Relative variation in organisms' infection risk 
as a function of the mobility restrictions trigger for various transmission and mortality rates. The red, green, yellow, and blue lines indicate the simulations for $\kappa=1.0$, $\kappa=2.0$, $\kappa=3.0$, and $\kappa=4.0$, respectively, in Fig.~\ref{fig9a}, and $\mu=0.1$, $\mu=0.2$, $\mu=0.3$, and $\mu=0.4$, respectively, in Fig.~\ref{fig9b}. The outcomes were averaged using sets of $100$ realisations for $\nu=0.9$; the error bars indicate the standard deviation. The simulations in Figs.~\ref{fig9a} and ~\ref{fig9b} ran for $\mu=0.2$ and $\kappa=2.0$, respectively.}
  \label{fig9}
\end{figure}
\begin{figure}[t]
 \centering
       \begin{subfigure}{.48\textwidth}
        \centering
        \includegraphics[width=85mm]{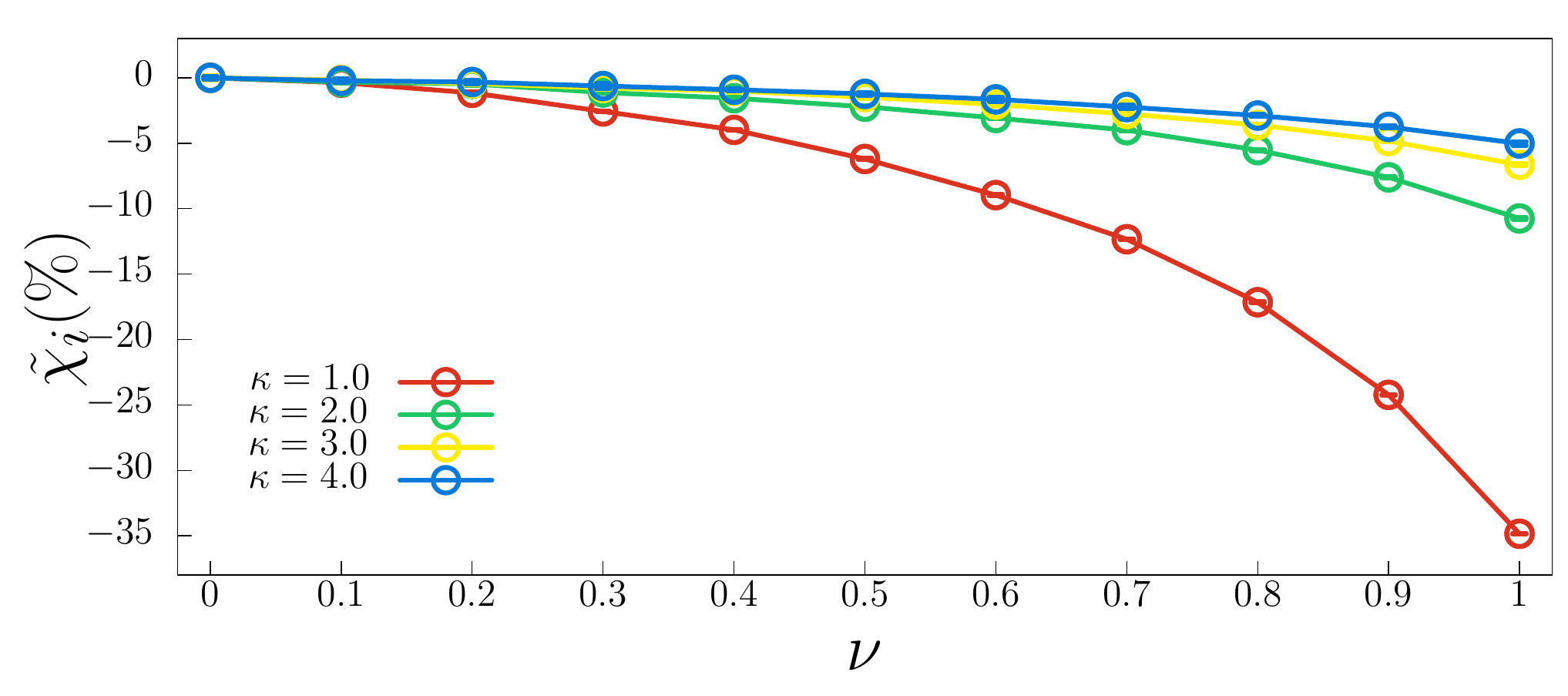}
        \caption{}\label{fig10a}
    \end{subfigure}
           \begin{subfigure}{.48\textwidth}
        \centering
       \includegraphics[width=85mm]{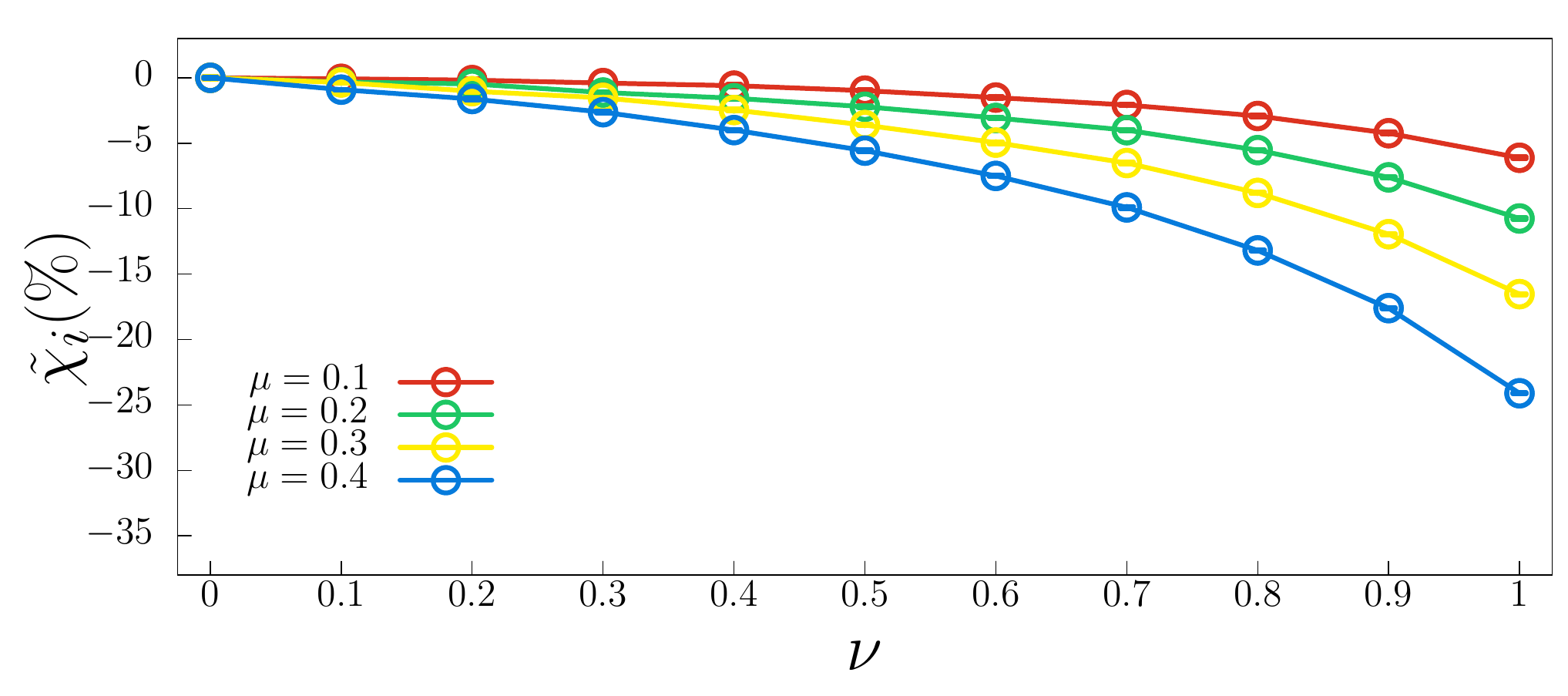}
        \caption{}\label{fig10b}
    \end{subfigure}
\caption{Relative variation in organisms' infection risk 
as a function of slowness factors for various transmission and mortality rates. The red, green, yellow, and blue lines depict the outcomes for $\kappa=1.0$, $\kappa=2.0$, $\kappa=3.0$, and $\kappa=4.0$, respectively, in Fig.~\ref{fig10a}, and $\mu=0.1$, $\mu=0.2$, $\mu=0.3$, and $\mu=0.4$, respectively, in Fig.~\ref{fig10b}. We performed sets of $100$ realisations for $\nu=0.9$; the error bars indicate the standard deviation.  The simulations in Figs.~\ref{fig10a} and ~\ref{fig10b} ran for $\mu=0.2$ and $\kappa=2.0$, respectively.}
  \label{fig10}
\end{figure}

The relative change in the infection risk of organisms of species $1$ is given by
\begin{equation}
\tilde{\chi_1}\,=\, \frac{\chi_1-\chi_1^0}{\chi_1^0},
\end{equation}
with $\chi_1^0$ being the infection risk of individuals of species $1$ in the absence of strategic mobility reduction, $\varphi=1.0$ and $\nu=0.0$.
We performed $100$ simulations for each scenario, starting from different initial conditions. 

Figures \ref{fig9a} and \ref{fig9b} show the average $\tilde{\chi_1}$ for $0 \leq \varphi \leq 1$, in intervals of $\Delta\varphi=0.1$, running for $\nu=0.9$.
Figures \ref{fig10a} and \ref{fig10b} depict the results for $0 \leq \nu \leq 1$, with $\Delta\nu=0.1$ and trigger $\varphi=0.1$.
In Figs. \ref{fig9a} and \ref{fig10a}, the red, green, yellow, and blue lines show the outcomes of the simulations for the mortality rate $\mu=0.2$, running 
with the transmission parameter $\kappa=1.0$, $\kappa=2.0$, $\kappa=3.0$, and $\kappa=4.0$, respectively. In addition, Figs. \ref{fig9b} and \ref{fig10b} depict $\tilde{\chi_1}$ for $\kappa=2.0$ with $\mu=0.1$ (red line), $\mu=0.2$ (green line), $\mu=0.3$ (yellow line), and $\mu=0.4$ (blue line).

We discover that the more contagious or less mortal the disease becomes when a pathogen mutation occurs, the less effective the mobility restriction response is in ensuring organisms' infection risk reduction. Even though the rise in the density of sick organisms leads 
more organisms to decelerate, disease transmission continues to be sustained. Conversely, if the disease outbreak is weak, concentrations of sick individuals are less spatially correlated, leading to the locally adaptive mobility restriction tactic of species $1$ to be efficient. For example, for $\kappa=1.0$, the reduction in the organisms' contamination risk is $25\%$, while for $\kappa=4.0$, the profit of the mobility restriction reaction drops to less than $4\%$.

In summary, our findings reveal that:
\begin{enumerate}
\item
Figure \ref{fig9} shows that if organisms of species $1$ trigger the mobility restriction tactic only if the local density of sick individuals exceeds $70\%$, the strategy is ineffective in protecting against contamination. In this scenario, the number of viral vectors surrounding the organisms is too high; thus, the survival strategy is not helpful. The reduction in the infection risk is optimised if $\varphi$ is lowered if the disease becomes less transmissible or less mortal.
\item
Regarding the mobility restriction level, Fig. \ref{fig10} shows that the mobility restriction response is more effective for accentuated limitations, irrespective of the disease virulence. For low $\kappa$ and high $\mu$, where the density of viral vectors in the system is low, a not strict mobility restriction can guarantee relevant benefits to the organisms' disease contamination prevention. 
\end{enumerate}

\section{Discussion and Conclusions}
\label{sec8}
We perform stochastic simulations of the spatial rock-paper-scissors models with individuals facing a contagious disease epidemic. The cyclic game is unbalanced by an evolutionary 
strategy performed by individuals of one out of the species to minimise the chances of contamination. Namely, each organism autonomously scans the vicinity, reducing its velocity whenever the fraction of viral vectors exceeds a tolerable threshold; otherwise, the individual continues exploring the territory moving with maximum speed. 

Our investigation considers a range of disease virulence, simulating diseases with many transmission and mortality rates. 
We discovered that mobility unevenness in cyclic spatial game induced by local disease outbreaks accentuates if a pathogen mutation produces more contagious or less mortal sicknesses. If organisms of species $i$ react to the arrival of a disease surge in the environment, species $i+1$ predominates in the spatial game, occupying departed spatial domains with the largest extension.

The simulations provide evidence that the speed reduction of individuals of species $i$ in response to local outbreaks
is effective in minimising the organisms' infection risk.
More, organisms of species $i-1$ also profit from the behaviour of individuals of species $i$. Due to the dominance of species $i+1$, when invading territories of species $i$, organisms conquer areas with a lower average density of sick individuals.
In contrast, individuals of species $i+1$ are the most likely to become sick because the deceleration of individuals of species $i$ delays the invasion rate. This benefits species $2$, whose population grows, but makes the disease spread more sustainably among individuals. 

The results show that mobility must be limited before the virus infects most neighbours. This means that the lower the local density of sick individuals is defined as the trigger, the more significant the protection achieved against contamination. 
In addition, the more rigid the movement constraint imposed by the individual on itself, the more the infection risk is reduced.
The outcomes also show that the mobility restriction strategy is more 
effective if the intensity of disease outbreaks is low, which happens in cases of low transmission or high mortality rates. This allows organisms to alternate between moving temporarily slowly and resuming the usual speed as soon as the local infection wave ends.

Our conclusions can be generalised for more complex systems where organisms can retain a temporary immunity after being cured of the disease - or if some individuals are vaccinated. If the inoculated organisms are no longer vulnerable to infection, they act as obstacles to disease propagation, reducing the average intensity of the local outbreaks; in this scenario, the mobility restriction reaction becomes more effective. Conversely, if the vaccine (or immunity by infection) only reduces death risk but does not protect against contamination, the number of ill individuals in the system rises; thus, individuals benefit less from the behavioural strategy.

Our model may provide insight to ecologists devoted to understanding the evolutionary behaviour in response to environmental changes threatening organisms' survival. Furthermore, our outcomes highlight the importance of the local
interactions and the consequences at individual and population levels, impacting ecosystem stability.
\section*{Acknowledgments}
We thank CNPq, ECT, Fapern, and IBED for financial and technical support.
\bibliographystyle{elsarticle-num}
\bibliography{ref}

\begin{thebibliography}{10}
\expandafter\ifx\csname url\endcsname\relax
  \def\url#1{\texttt{#1}}\fi
\expandafter\ifx\csname urlprefix\endcsname\relax\def\urlprefix{URL }\fi
\expandafter\ifx\csname href\endcsname\relax
  \def\href#1#2{#2} \def\path#1{#1}\fi

\bibitem{bacteria}
B.~C. Kirkup, M.~A. Riley, Antibiotic-mediated antagonism leads to a bacterial
  game of rock-paper-scissors in vivo, Nature 428 (2004) 412--414.

\bibitem{Coli}
B.~Kerr, M.~A. Riley, M.~W. Feldman, B.~J.~M. Bohannan, Local dispersal
  promotes biodiversity in a real-life game of rock–paper–scissors, Nature
  418 (2002) 171.

\bibitem{Allelopathy}
R.~Durret, S.~Levin, Allelopathy in spatially distributed populations, J.
  Theor. Biol. 185 (1997) 165--171.

\bibitem{ecology}
M.~Begon, C.~R. Townsend, J.~L. Harper, Ecology: from individuals to
  ecosystems, Blackwell Publishing, Oxford, 2006.

\bibitem{Nature-bio}
A.~Purvis, A.~Hector, Getting the measure of biodiversity, Nature 405 (2000)
  212--2019.

\bibitem{mobilia2}
T.~Reichenbach, M.~Mobilia, E.~Frey, Mobility promotes and jeopardizes
  biodiversity in rock-paper-scissors games, Nature 448 (2007) 1046--1049.

\bibitem{mobilia3}
T.~Reichenbach, M.~Mobilia, E.~Frey, Self-organization of mobile populations in
  cyclic competition, Journal of Theoretical Biology 254~(2) (2008) 368 -- 383.

\bibitem{lizards}
B.~Sinervo, C.~M. Lively, The rock-scissors-paper game and the evolution of
  alternative male strategies, Nature 380 (1996) 240--243.

\bibitem{coral}
J.~B.~C. Jackson, L.~Buss, The rock-scissors-paper game and the evolution of
  alternative male strategies, Proc. Natl Acad. Sci. USA 72 (1975) 5160--5163.

\bibitem{mobiliahigh}
S.~Islam, A.~Mondal, M.~Mobilia, S.~Bhattacharyya, C.~Hens, Effect of mobility
  in the rock-paper-scissor dynamics with high mortality, Phys. Rev. E 105
  (2022) 014215.

\bibitem{foraging}
P.~A. Abrams, Foraging time optimization and interactions in food webs, The
  American Naturalist 124~(1) (1984) 80--96.

\bibitem{butterfly}
A.~Cormont, A.~H. Malinowska, O.~Kostenko, V.~Radchuk, L.~Hemerik, M.~F.
  WallisDeVries, J.~Verboom, Effect of local weather on butterfly flight
  behaviour, movement, and colonization: significance for dispersal under
  climate change, Biodiversity and Conservation 20 (2011) 483--503.

\bibitem{BUCHHOLZ2007401}
R.~Buchholz, Behavioural biology: an effective and relevant conservation tool,
  Trends in Ecology \& Evolution 22~(8) (2007) 401 -- 407.

\bibitem{doi:10.1002/ece3.4446}
A.~M. Revynthi, M.~Egas, A.~Janssen, M.~W. Sabelis, Prey exploitation and
  dispersal strategies vary among natural populations of a predatory mite,
  Ecology and Evolution 8~(21) (2018) 10384--10394.

\bibitem{adaptive1}
L.~Riotte-Lambert, J.~Matthiopoulos, Environmental predictability as a cause
  and consequence of animal movement, Trends in Ecology \& Evolution 35~(2)
  (2020) 163--174.

\bibitem{adaptive2}
P.~A. Abrams, Habitat choice in predator‐prey systems: Spatial instability
  due to interacting adaptive movements, The American Naturalist 169~(5) (2007)
  581--594.

\bibitem{Dispersal}
D.~Bonte, M.~Dahirel, Dispersal: a central and independent trait in life
  history, Oikos 126 (2017) 472--479.

\bibitem{BENHAMOU1989375}
S.~Benhamou, P.~Pierre~Bovet, How animals use their environment: a new look at
  kinesis, Animal Behaviour 38~(3) (1989) 375--383.

\bibitem{Moura}
B.~Moura, J.~Menezes, Behavioural movement strategies in cyclic models,
  Scientific Reports 11 (2021) 6413.

\bibitem{anti2}
J.~Menezes, B.~Moura, Mobility-limiting antipredator response in the
  rock-paper-scissors model, Phys. Rev. E 104 (2021) 054201.

\bibitem{MENEZES2022101606}
J.~Menezes, E.~Rangel, B.~Moura, Aggregation as an antipredator strategy in the
  rock-paper-scissors model, Ecological Informatics 69 (2022) 101606.

\bibitem{TENORIO2022112430}
M.~Tenorio, E.~Rangel, J.~Menezes, Adaptive movement strategy in
  rock-paper-scissors models, Chaos, Solitons \& Fractals 162 (2022) 112430.

\bibitem{Menezes_2022}
J.~Menezes, M.~Tenorio, E.~Rangel, Adaptive movement strategy may promote
  biodiversity in the rock-paper-scissors model, Europhysics Letters 139~(5)
  (2022) 57002.

\bibitem{animats}
P.~Maes, M.~J. Mataric, J.~A. Meyer, J.~Pollack, S.~W. Wilson, Some adaptive
  movements of animats with single symmetrical sensors, 1996, pp. 55--64.

\bibitem{social1}
E.~Du, E.~Chen, J.~Liu, C.~Zheng, How do social media and individual behaviors
  affect epidemic transmission and control?, Science of The Total Environment
  761 (2021) 144114.

\bibitem{disease4}
A.~M. Dunn, M.~E. Torchin, M.~J. Hatcher, P.~M. Kotanen, D.~M. Blumenthal,
  J.~E. Byers, C.~A. Coon, V.~M. Frankel, R.~D. Holt, R.~A. Hufbauer, A.~R.
  Kanarek, K.~A. Schierenbeck, L.~M. Wolfe, S.~E. Perkins, Indirect effects of
  parasites in invasions, Functional Ecology 26~(6) (2012) 1262--1274.

\bibitem{tanimoto}
J.~Tanimoto, Sociophysics Approach to Epidemics, Springer,, Singapore, 2021.

\bibitem{socialdist}
T.~C. Reluga, Game theory of social distancing in response to an epidemic, PLoS
  Comput. Biol. 6~(5) (2010) e1000793.

\bibitem{soc}
S.~Stockmaier, N.~Stroeymeyt, S.~E. C., H.~D. M., L.~A. Meyers, D.~I. Bolnick,
  Infectious diseases and social distancing in nature, Science 371~(6533)
  (2021) eabc8881.

\bibitem{doi:10.1126/science.abc8881}
S.~Stockmaier, N.~Stroeymeyt, E.~C. Shattuck, D.~M. Hawley, L.~A. Meyers, D.~I.
  Bolnick, Infectious diseases and social distancing in nature, Science
  371~(6533)  eabc8881.

\bibitem{mr0}
G.~Dimarco, G.~Toscani, M.~Zanella, Optimal control of epidemic spreading in
  the presence of social heterogeneity, Philosophical Transactions of the Royal
  Society A: Mathematical, Physical and Engineering Sciences 380~(2224) (2022)
  20210160.

\bibitem{mr1}
Q.~Shao, D.~Han, Epidemic spreading in metapopulation networks with
  heterogeneous mobility rates, Applied Mathematics and Computation 412 (2022)
  126559.

\bibitem{mr2}
P.~Edsberg~Møllgaard, S.~Lehmann, L.~Alessandretti, Understanding components
  of mobility during the covid-19 pandemic, Philosophical Transactions of the
  Royal Society A: Mathematical, Physical and Engineering Sciences 380~(2214)
  (2022) 20210118.

\bibitem{10.1371/journal.pone.0254403}
T.~Oka, W.~Wei, D.~Zhu, The effect of human mobility restrictions on the
  covid-19 transmission network in china, PLOS ONE 16~(7) (2021) 1--16.

\bibitem{CAPAROGLU2021111246}
To restrict or not to restrict? use of artificial neural network to evaluate
  the effectiveness of mitigation policies: A case study of turkey, Chaos,
  Solitons \& Fractals 151 (2021) 111246.

\bibitem{plasticity2}
N.~Stroeymeyt, A.~V. Grasse, A.~Crespi, D.~P. Mersch, S.~Cremer, L.~Keller,
  Social network plasticity decreases disease transmission in a eusocial
  insect, Science 362~(6417) (2018) 941--945.

\bibitem{Gene}
V.~Papanikolaou, A.~Chrysovergis, V.~Ragos, E.~Tsiambas, S.~Katsinis,
  A.~Manoli, S.~Papouliakos, D.~Roukas, S.~Mastronikolis, D.~Peschos,
  A.~Batistatou, E.~Kyrodimos, N.~Mastronikolis, From delta to omicron:
  S1-rbd/s2 mutation/deletion equilibrium in sars-cov-2 defined variants, Gene
  814 (2022) 146134.

\bibitem{mutating1}
M.~Becerra-Flores, T.~Cardozo, Sars-cov-2 viral spike g614 mutation exhibits
  higher case fatality rate, International Journal of Clinical Practice 74~(8)
  (2020) e13525.

\bibitem{mutate2}
H.~M. Zawbaa, H.~Osama, A.~El-Gendy, H.~Saeed, H.~S. Harb, Y.~M. Madney,
  M.~Abdelrahman, M.~Mohsen, A.~M.~A. Ali, M.~Nicola, M.~O. Elgendy, I.~A.
  Ibrahim, M.~E.~A. Abdelrahim, Effect of mutation and vaccination on spread,
  severity, and mortality of covid-19 disease, Journal of Medical Virology
  94~(1) (2022) 197--204.

\bibitem{plasticity1}
A.~Bridier, P.~C. Piard, J-C.~and, S.~Labarthe, F.~Dubois-Brissonnet,
  R.~Briandet, Spatial organization plasticity as an adaptive driver of surface
  microbial communities, Front. Microbiol 8 (2017) 1364.

\bibitem{plasticity}
Spatial organisation plasticity reduces disease infection risk in
  rock–paper–scissors models, Biosystems 221 (2022) 104777.

\bibitem{PhysRevE.97.032415}
P.~P. Avelino, D.~Bazeia, L.~Losano, J.~Menezes, B.~F. de~Oliveira, M.~A.
  Santos, How directional mobility affects coexistence in rock-paper-scissors
  models, Phys. Rev. E 97 (2018) 032415.

\bibitem{Avelino-PRE-86-036112}
P.~P. Avelino, D.~Bazeia, L.~Losano, J.~Menezes, B.~F. Oliveira, Junctions and
  spiral patterns in generalized rock-paper-scissors models, Phys. Rev. E 86
  (2012) 036112.

\bibitem{BAZEIA2022126547}
D.~Bazeia, M.~Bongestab, B.~{de Oliveira}, Influence of the neighborhood on
  cyclic models of biodiversity, Physica A: Statistical Mechanics and its
  Applications 587 (2022) 126547.

\bibitem{combination}
E.~Rangel, B.~Moura, J.~Menezes, Combination of survival movement strategies in
  cyclic game systems during an epidemic, Biosystems 217 (2022) 104689.

\bibitem{jcomp}
J.~Menezes, B.~Moura, E.~Rangel, Adaptive survival movement strategy to local
  epidemic outbreaks in cyclic models, Journal of Physics: Complexity 3~(4)
  (2022) 045008.

\bibitem{eplsick}
J.~Menezes, B.~Ferreira, E.~Rangel, B.~Moura, Adaptive altruistic strategy in
  cyclic models during an epidemic, Europhysics Letters 140~(5) (2022) 57001.

\bibitem{rps-epidemy}
W.-X. Wang, Y.-C. Lai, C.~Grebogi, Effect of epidemic spreading on species
  coexistence in spatial rock-paper-scissors games, Phys. Rev. E 81 (2010)
  046113.

\bibitem{epidemic-graphs}
T.~Nagatani, Infection promotes species coexistence: Rock–paper–scissors
  game with epidemic on graphs, Physica A: Statistical Mechanics and its
  Applications 535 (2019) 122531.

\bibitem{germen}
X.~Yang, H.~Zhou, X.~Zhou, Rock paper scissors: Crispr/cas9-mediated
  interference with geminiviruses in plants, Science China Life Sciences (2019)
  1389--1391.

\bibitem{unevenmobility}
J.~Menezes, S.~Rodrigues, S.~Batista, Mobility unevenness in
  rock–paper–scissors models, Ecological Complexity (2022) 101028.

\bibitem{uneven}
J.~Menezes, B.~Moura, T.~A. Pereira, Uneven rock-paper-scissors models:
  Patterns and coexistence, Europhysics Letters 126~(1) (2019) 18003.

\bibitem{PedroWeak}
P.~P. Avelino, B.~F. de~Oliveira, R.~S. Trintin, Predominance of the weakest
  species in lotka-volterra and may-leonard formulations of the
  rock-paper-scissors model, Phys. Rev. E 100 (2019) 042209.

\bibitem{unevenpark1}
J.~Park, Balancedness among competitions for biodiversity in the cyclic
  structured three species system, Applied Mathematics and Computation 320
  (2018) 425--436.

\bibitem{howlocal}
J.~Menezes, S.~Batista, M.~Tenorio, E.~Triaca, B.~Moura, How local antipredator
  response unbalances the rock-paper-scissors model, Chaos: An
  Interdisciplinary Journal of Nonlinear Science 32~(12) (2022) 123142.

\bibitem{MENEZES2023113290}
J.~Menezes, R.~Barbalho, How multiple weak species jeopardise biodiversity in
  spatial rock–paper–scissors models, Chaos, Solitons \& Fractals 169
  (2023) 113290.

\bibitem{leonard}
R.~M. May, W.~J. Leonard, Nonlinear aspects of competition between three
  species, SIAM J. Appl. Math. 29 (1975) 243--253.

\bibitem{random}
S.~Redner, A Guide to First-Passage Processes, Cambridge Univ. Press, Cambridge
  Univ. Press, 2001.

\end{thebibliography}

\end{document}